\documentclass[journal]{IEEEtran}
\usepackage{graphicx}
\usepackage{epstopdf}
\usepackage{cite}
\usepackage{amsfonts}
\usepackage{amssymb}
\usepackage{amsmath}
\usepackage{mathrsfs}
\usepackage{bm}
\usepackage{amsmath}
\usepackage{mdwmath}
\usepackage{hyperref}
\usepackage{mdwtab}
\usepackage[ruled,vlined,linesnumbered]{algorithm2e} 

\newcommand\smallO{
	\mathchoice
	{{\scriptstyle\mathcal{O}}}
	{{\scriptstyle\mathcal{O}}}
	{{\scriptscriptstyle\mathcal{O}}}
	{\scalebox{.7}{$\scriptscriptstyle\mathcal{O}$}}
}

\newtheorem{remark}{Remark}
\newtheorem{corollary}{Corollary}
\newtheorem{theorem}{Theorem}
\newtheorem{prop}{Proposition}
\DeclareMathOperator*{\argmax}{arg\,max}

\begin{document}

\title{Optimal Discrete Spatial Compression for Beamspace Massive MIMO Signals}
\author{Zhiyuan Jiang, Sheng Zhou, Zhisheng Niu,~\IEEEmembership{Fellow,~IEEE}
\thanks{Z. Jiang, S. Zhou and Z. Niu are with Tsinghua National Laboratory for Information Science and Technology, Tsinghua University, Beijing 100084, China. Emails: \{zhiyuan, sheng.zhou, niuzhs\}@tsinghua.edu.cn. This work is sponsored in part by the Nature Science Foundation of China (No. 61701275, No. 91638204, No. 61571265), China Postdoctoral Science Foundation and Intel Collaborative Research Institute for Mobile Networking and Computing. The corresponding author is Sheng Zhou. Part of the work has been submitted to IEEE International Conference of Communications 2018.}}
\maketitle

\begin{abstract}
	Deploying massive number of antennas at the base station side can boost the cellular system performance dramatically. Meanwhile, it however involves significant additional radio-frequency (RF) front-end complexity, hardware cost and power consumption. To address this issue, the beamspace-multiple-input-multiple-output (beamspace-MIMO) based approach is considered as a promising solution. In this paper, we first show that the traditional beamspace-MIMO suffers from spatial power leakage and imperfect channel statistics estimation. A beam combination module is hence proposed, which consists of a small number (compared with the number of antenna elements) of low-resolution (possibly one-bit) digital (discrete) phase shifters after the beamspace transformation to further compress the beamspace signal dimensionality, such that the number of RF chains can be reduced beyond beamspace transformation and beam selection. The optimum discrete beam combination weights for the uplink are obtained based on the branch-and-bound (BB) approach. The key to the BB-based solution is to solve the embodied sub-problem, whose solution is derived in a closed-form. Based on the solution, a sequential greedy beam combination scheme with linear-complexity (w.r.t. the number of beams in the beamspace) is proposed. Link-level simulation results based on realistic channel models and long-term-evolution (LTE) parameters are presented which show that the proposed schemes can reduce the number of RF chains by up to $25\%$ with a one-bit digital phase-shifter-network. 
\end{abstract}


\section{Introduction}
Multi-antenna technology, or multiple-input multiple-output (MIMO) systems will play a pivotal role in the next-generation cellular systems. Deploying a large number of antennas (massive MIMO) at the base station (BS) side brings tremendous spatial degree-of-freedoms (DoFs), which can significantly improve the system performance. Specifically, the benefits of massive MIMO systems include improved signal coverage which is instrumental for millimeter-wave systems, better interference management and hence larger spatial multiplexing gain, and also enhanced link reliability for critical machine type communication (C-MTC) applications. Therefore, the multi-antenna technology is a key enabler for various promising applications in the $5$G cellular systems. 

A large body of researches have been dedicated to studying the massive MIMO system performance under the assumption of full digital baseband signal processing \cite{Marzetta10,Rusek12,larsson14}. It is required thereby that each antenna element has one dedicated radio-frequency (RF) chain associated with it. Existing work shows that significant spectral and radiated energy efficiency improvements can be achieved by full digital processing. However, it is also widely recognized that digitally controlling all antenna elements poses severe challenges, making it infeasible to realize. First, full digital signal processing is expensive, both in terms of \textbf{hardware cost} and \textbf{power consumption}. Full digital signal processing requires that one RF chain, including e.g., low-noise amplifier, analog-digital-converter (ADC), power amplifier and etc., is needed for each antenna element. As the number of antenna elements is scaled up in massive MIMO systems, this requirement entails a dramatic increase in the deployment cost of the system. Moreover, the power consumption would also be driven up to a prohibitive level. As indicated in the existing work \cite{heath16} \cite{han15}, concretely, a BS with $256$ RF chains, which is considered a moderate number in massive MIMO systems, consumes (only the RF chains) about $10$ fold the power of an entire current long-term-evolution (LTE) BS. On the other hand, full digital signal processing entails \textbf{enormous computational and pilot overhead}. Spatial baseband processing includes multiple matrix operations, such as inversions and singular-value-decomposition (SVD) whose complexity scales with $M^3$ where $M$ is the number of antenna elements. In addition, these extremely computational demanding matrix operations are required to be executed very frequently (once every $1$~{ms} for spatial precoding in LTE systems). Besides, the channel state information (CSI) acquisition overhead, i.e., pilot overhead, in frequency-division-duplexing (FDD) system scales with the number of digitally controllable antennas which equals with the number of RF chains. It constitutes a major bottleneck in realizing the massive MIMO gain in FDD systems \cite{xie16,Jiang14}. 

\begin{figure*}[!t]
	\centering
	\includegraphics[width=0.95\textwidth]{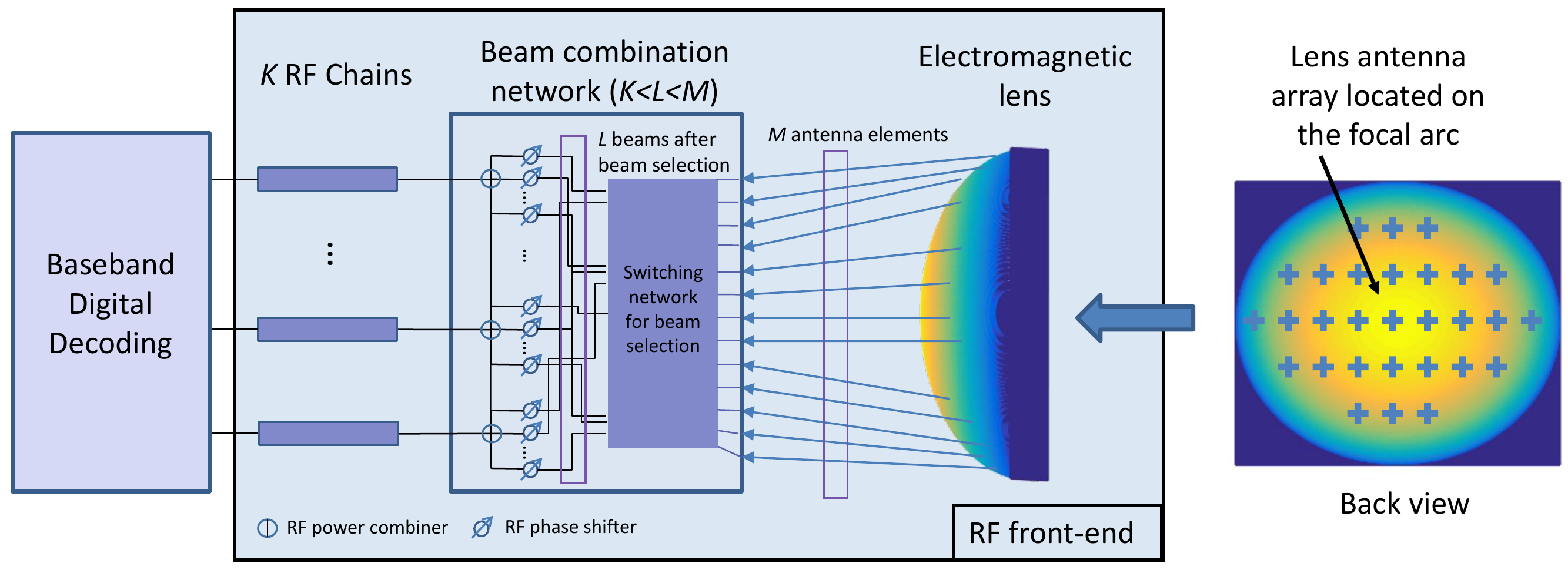}
	\caption{Proposed beamspace massive MIMO system overview. The phase shifter network is used for beam combination to further reduce the RF complexity, consisting of finite-resolution (possibly one-bit) digital phase shifters.}
	\label{Fig_arch}
\end{figure*}

In view of these challenges, the hybrid beamforming architecture has been proposed \cite{zhang05, alkh14,molisch16,brady13,Jiang14,jiang17}. It adopts an RF analog beamforming module to generate beams. The RF chains are attached to beams instead of antennas, and hence the number of RF chains is significantly reduced due to angular power concentration \cite{molisch16}. However, such systems suffer from the cost issues since the analog beamforming module consists of a large number (scales with the number of antenna elements) of phase shifters to adjust signal directions. The cost is even higher in millimeter-wave systems due to the high frequency range which requires better RF circuit quality, e.g., stray capacitances and circuit Q factor. Some work proposes to replace phase shifters with switches, i.e., performing antenna selection, but with notable performance degradation \cite{molisch04mag,zhang05}. A promising solution is the lens antenna array, or beamspace MIMO architecture \cite{dus11,brady13,gao16_bs,ama15,zeng16,zeng17}, by which the analog beamforming module is a lens antenna array. See Fig. \ref{Fig_arch} for an instance. By analogy, the lens antenna array focuses on each direction of the incoming (or outgoing) electromagnetic wave, just as a focal lens on beams of visible light. In this way, the signals are transformed to the angular domain (beam domain), such that each angular bin (beam) only contains the signal from a specific direction. Mathematically, assuming one-dimensional array, the lens antenna array performs a discrete-Fourier-transform (DFT) to the antenna domain signals. The DFT length equals the number of antennas\footnote{The two-dimensional (planar array) transformation is the Kronecker product of two one-dimensional DFT.}. This is achieved without any phase shifters or beamforming codebook design. The key reason that beamspace MIMO can reduce the number of RF chains is the angular power sparsity discovered in massive MIMO channels, especially for millimeter-wave systems \cite{bas17}\cite{rap13}. Therefore, some subset of the beams contains nearly all the signal power. Accordingly, the RF chains are only attached to the selected beams. Such an approach is shown to be very effective which can reduce the number of RF chains dramatically with little performance degradation \cite{ama15}. 

In this paper, in addition to the beamspace massive MIMO transformation, we aim to further reduce the number of RF chains by combining different correlated beams with a low-cost phase shifter network (PSN). The architecture is described in Fig. \ref{Fig_arch}. Since the RF beamforming module should be simplified with low-cost to enable wide usage in, e.g., remote-radio-units (RRUs) in cloud radio access networks (C-RAN) systems, the beam combination module is composed of low-resolution ($B$ bits which equals $2^B$ phase shifting states and possibly one-bit for $B=1$) digital phase shifters with constant amplitudes \cite{ven10}.
The main contributions include:
\begin{itemize}
	\item 
	First, we show by examples that the current beamspace MIMO transformation has potential to be improved regarding RF chain reduction, due to spatial power leakage and imperfect channel statistics estimation. 
	\item
	The optimal beam combination with arbitrary weights is derived for the uplink, which is related to the dominant signal eigenspace of the signal. The achieved performance serves as an upper bound benchmark for hardware-constrained combination methods. In view of the hardware constrains, a branch-and-bound (BB) method is proposed to obtain the optimum discrete combining weights with unit-amplitude and limited-phase-resolution digital phase shifters. The most prominent contribution is that we provide a closed-form solution for the sub-problem in the BB method. This enables us to propose a sequential greedy beam combination (SG-BC) which shows near-optimal performance with significantly reduced complexity. 
	\item
	We conduct realistic link-level simulations with canonical 3rd Generation Partnership Project (3GPP) spatial channel models to validate our proposed schemes. To encourage reproducibility, the simulation MATLAB codes are available at \url{https://github.com/battleq2q/Beam-Combination-For-Beamspace-MIMO.git}.
\end{itemize}

The remainder of the paper is organized as follow. In Section \ref{sec:model}, the system model and the proposed system architecture are introduced. Preliminaries about the array response of beamspace MIMO is also illustrated. In Section \ref{sec:me}, the reason why beamspace transformation is not enough to reduce RF complexity is illustrated by examples. In Section \ref{sec:st}, the optimal spatial compression schemes are described and the optimality proofs are given. Simulation results with practical channel models and LTE numerologies are presented in Section \ref{sec:sr}. Finally in Section \ref{sec:conc}, we conclude the work and discuss some future directions. 

Throughout the paper, we use boldface uppercase letters, boldface lowercase letters and lowercase letters to designate matrices, column vectors and scalars, respectively. The symbol $j$ represents the imaginary unit of complex numbers, with $j^2 = -1$. $\bm{X}^\dag$ denotes the complex conjugate transpose of matrix $\bm{X}$. $[\bm{X}]_{i,j}$ and $x_i$ denotes the $(i,j)$-th entry and $i$-th element of matrix $\bm{X}$ and vector $\bm{x}$, respectively. $\textnormal{tr}(\bm{X})$ denotes the trace of matrix $\bm{X}$. Denote by $\bm{A} \otimes \bm{B}$ as the Kronecker product of $\bm{A}$ and $\bm{B}$. The Euclidean norm of a vector $\bm{x}$ is denoted by $\|\bm{x}\|_2$. The vectorization of a matrix $\bm{X}$, denoted by vec($\bm{X}$), is the column vector obtained by stacking the columns of the matrix $\bm{X}$. Denote by $\mathbb{E}(\cdot)$ as the expectation operation. Denote by $\bm{I}_N$ as the $N$ dimensional identity matrix. Denote by $\bm{0}_N$ as a $N$-dimensional zero vector. The logarithm $\log(x)$ denotes the binary logarithm. The phase of a complex-valued number $x$ is denoted by $\angle x$. An empty set, as well as an empty matrix, is denoted by $\phi$.

\subsection{Related Work}
The proposed spatial compression schemes are related to the subspace tracking methods proposed in, e.g., \cite{sae15,liu15,zhou16,jiang_icc17}. These works propose the full digital (mostly SVD-based) spatial compression to enhance channel estimation performance, or reduce the fronthaul interface transmission rate in C-RAN. However, the results are not applicable to limited-resolution PSN-based beam combination. Methods that compress the signal in other domains such as frequency and time domains are studied in, e.g., \cite{Chen11}\cite{guo12}. In beamspace MIMO systems, there is very little work on reducing the RF complexity beyond the beamspace transformation and beam selection, even though the DFT power leakage problem is pointed out in \cite{gao16_bs}\cite{zeng16}. However, only the ideal multi-path propagation environment is considered, where the spatial power leakage is ignored.

Regarding the beamforming methods with hardware limitation \cite{lan17,gok17,mo15,sax16,jacob16,landau13,bj15,israel13,fet07,ven10}, the work in \cite{fet07,bj15} proposes to use dirty RF in massive MIMO systems. The dirty RF design philosophy is to use RF hardware that is with low-cost and low precision, e.g., $1$-bit ADCs, thanks to the excess DoFs to counteract the hardware imperfections. However, most work focuses on the limited-resolution ADC and digital-analog-converter (DAC) designs. The work in \cite{ven10}\cite{israel13} considers the discrete beamforming design with limited-resolution PSNs. However, it considers the Capon method wherein the objective is to mitigate the multi-path interference, and is entirely different from this work. Authors in \cite{sax16}\cite{jacob16} adopt the sub-optimal consecutive quantization of linear precoding strategies. In \cite{lan17} \cite{landau13}, the branch-and-bound method is adopted whereas the sub-problem is not solved in a closed-form and a sub-optimal approach is adopted.

\section{System Model and Preliminaries}
\label{sec:model}
\subsection{Signal Model}
We consider the uplink (UL) of a single cell system. The UL baseband equivalent signal model before going through the lens antenna array is written as
\begin{equation}
\label{hx}
\bm{y} = \bm{H}\bm{x}+\bm{n},
\end{equation}
where the UL receive signal, i.e., $\bm{y}$, is a complex vector of dimension $M$, and $M$ is the total number of antenna elements. Vector $\bm{x}$ is the uplink transmit signals from $N$ users. The equivalent identically-independently distributed (i.i.d.) Gaussian additive noise is denoted by $\bm{n}$ which is added here for ease of exposition. Denote by $\bm{h}_{n,i}$ as the channel vector from the $i$-th antenna of user $n$ to $M$  receive antennas. Each user is equipped with $A_n$ antennas\footnote{In this paper, uplink transmit beamforming is not considered although users are assumed to be equipped with multiple antennas. However, it is reasonable to conclude that the angular power spectrum at the BS side would be even more sparse when uplink directional beamforming is implemented. Hence, the spatial compression gain should increase accordingly.}, and together they form $\bm{x}$ of dimension $A=\sum\limits_{n=1}^{N} A_n$. Denote by $\bm{H}$ is the channel matrix of dimension $M \times A$. Without loss of generality, the narrow-band signal model is adopted whereby the signal bandwidth is much smaller than the carrier frequency. Denote the signal after receive beamforming as
\begin{equation}
\label{channel_model}
\bm{c} = \bm{D}\bm{A}_\mathcal{C}\bm{A}_\mathcal{L}\bm{y},
\end{equation}
where the RF beamforming (beamspace transformation) at the lens antenna array and the beam selection are denoted by $\bm{A}_\mathcal{L} \in \mathbb{C}^{L \times M}$ ($L$ beams are selected), the beam combination proposed by this paper is denoted by $\bm{A}_\mathcal{C} \in \mathbb{C}^{K \times L} $, and the baseband digital receive beamforming is denoted by $\bm{D} \in \mathbb{C}^{A \times K}$ and hence $K$ is the number of RF chains. In this paper, the beam combination is subject to hardware constraints to reduce the RF hardware complexity. Therefore, the beam combination matrix is restricted to have unit-amplitude entries and limited phase-resolution, i.e.,
\begin{IEEEeqnarray}{rCl}
\label{ac_const}
&& \left[\bm{A}_\mathcal{C}\right]_{i,j} \in \bm{\Psi}, \, \bm{\Psi} = \left\{e^{\frac{j2n\pi}{2^B}},\,n = 0,...,2^B-1.\right\},
\end{IEEEeqnarray}
where $B$ is the resolution of the digital phase shifters, and $B=1$ denotes the one-bit PSN where there are only two states of the phase shifters, i.e., $\left[\bm{A}_\mathcal{C}\right]_{i,j} \in \{-1,1\}$.

\subsection{Channel Model}
Using a geometry-based channel model \cite{Molisch04}, the channel vector can be written as
\begin{equation}
\label{ray_rep}
\bm{h}_{n,i} = \sqrt{\frac{M}{U_{n,i}}}\sum\limits_{r=1}^{U_{n,i}}\beta_{n,r} \bm{\alpha}(\theta_{n,r},\psi_{n,r}),
\end{equation}
where $U_{n,i}$ denotes the total number of multi-path components (MPCs) in the propagation channel for the $i$-th antenna of user $n$, the amplitude of each MPC is denoted by $\beta_{n,r}$, and $\mathbb{E}[|\beta_{n,r}|^2]=\gamma_{n,r}$. The azimuth and elevation angle-of-arrival (AoA) of the $r$-th arriving MPC of user $n$ are denoted by $\theta_{n,r}$ and $\psi_{n,r}$, respectively. The steering vector for one MPC (assuming uniform linear array) is
\begin{IEEEeqnarray}{rCl}
\label{array_res}
&& \bm{\alpha}(\theta_{n,r}) = \frac{1}{\sqrt{M}}\left[e^{-j2\pi m\frac{d \sin{\theta_{n,r}}}{\lambda}}\right],\nonumber\\
&& m \in \left\{s-\frac{M-1}{2},s=0,1,...,M-1\right\}.
\end{IEEEeqnarray}
where $d$ is the antenna spacing\footnote{We assume the so-called critical antenna spacing, i.e., $d=\lambda/2$.} and $\lambda$ is the wavelength. Summing up all the contributing MPCs obtains the compound channel representation in \eqref{ray_rep}. For a judiciously designed lens antenna array, the beamspace transformation is equivalent to a DFT where each column of the DFT matrix is the uniform linear array signal from a specific AoA \cite{zeng16}, i.e., assuming without beam selection,
\begin{IEEEeqnarray}{rCl}
	\label{beamspace_tr}
\bm{A}_\mathcal{L} &=& \left[\bm{\alpha}(\theta_1),\bm{\alpha}(\theta_2),...,\bm{\alpha}(\theta_M)\right]^\dag, \nonumber\\
\sin{\theta_i}&=&\frac{\lambda}{d M}\left(i-\frac{M+1}{2}\right)
\end{IEEEeqnarray}
For a two-dimension lens antenna array, it can be derived that
\begin{equation}
\bm{A}_{\mathcal{L}, \textrm{2D}} =  \bm{A}_{\mathcal{L}, \textrm{row}} \otimes \bm{A}_{\mathcal{L}, \textrm{col}},
\end{equation}
The channel correlation matrix (CCM) of $\bm{h}_{n,i}$ is defined as 
\begin{equation}
\label{channel_corr}
\bm{R}_{n,i} = \mathbb{E}\left[\bm{h}_{n,i}\bm{h}_{n,i}^\dag\right].
\end{equation}
Denote the SVD of the CCM as
\begin{equation}
\label{r_svd}
\bm{R}_{n,i} = \bm{U}_{n,i}\bm{\Sigma}_{n,i}\bm{U}_{n,i}^\dag,
\end{equation}
where we always assume the singular values are arranged in non-increasing order. Combining \eqref{array_res} and \eqref{channel_corr} wherein the expectation is taken over MPC channel gain $\beta_{n,r}$, it follows that\footnote{It is assumed that the number of MPCs and the AoA of each MPC are stationary when estimating the CCM. This assumption is justified by the fact that the scattering statistics, including e.g., MPCs and AoAs, is relatively more static \cite{music} compared with channel gains $\beta_{n,r}$ and hence it can be assumed static in the given time period wherein the channel gains are averaged. }
\begin{equation}
\label{r_alpha}
\bm{R}_{n,i} = \frac{M}{U_{n,i}}\sum\limits_{r=1}^{U_{n,i}} \gamma_{n,r} \bm{\alpha}(\theta_{n,r},\psi_{n,r})\bm{\alpha}(\theta_{n,r},\psi_{n,r})^\dag.
\end{equation}
It is observed that the CCM is the summation of all the rank-$1$ matrices constructed by the steering vectors of MPCs. The cross terms of the steering vectors are averaged out because different MPCs usually have independently distributed small scale fading amplitude coefficients \cite{music}. It is straightforward to derive that
\begin{IEEEeqnarray}{rCl}
\label{en_CCM}
\bm{R}_\textrm{t} &=& \mathbb{E}\left[\bm{H}\bm{H}^\dag\right]=\mathbb{E}\left[\sum\limits_{n,i} \bm{h}_{n,i}\bm{h}_{n,i}^\dag\right] = \sum\limits_{n,i}  \bm{R}_{n,i} \nonumber\\
&=& \sum\limits_{n,i}\frac{M}{U_{n,i}}\sum\limits_{r=1}^{U_{n,i}} \gamma_{n,r} \bm{\alpha}(\theta_{n,r},\psi_{n,r})\bm{\alpha}(\theta_{n,r},\psi_{n,r})^\dag.
\end{IEEEeqnarray}
Alternatively, $\bm{R}_\textrm{t}$ can be interpreted as the overall CCM which consists of the MPCs from all users. The overall CCM is usually obtained by averaging the receive signal over a number of time and frequency resources, i.e.,
\begin{equation}
\label{ccm_cal}
\bar{\bm{R}}_\textrm{t} = \frac{1}{TL}\sum_{t,l}\bm{y}_{t,l}\bm{y}_{t,l}^\dag,
\end{equation}
where the average is over time (indexed by $t$) and frequency (indexed by $l$) receive symbols. The method is widely used in practice. More sophisticated CCM estimation algorithm can be found in, e.g., \cite{liang01} \cite{bickel08}. The time-averaged useful signal CCM is defined as
\begin{equation}
\label{ccm_u}
\bar{\bm{R}}_\textrm{s} = \frac{1}{TL}\sum_{t,l} \bm{H}_{t,l}\bm{x}_{t,l}\bm{x}_{t,l}^\dag \bm{H}_{t,l}^\dag,
\end{equation}
where
\begin{equation}
\label{R_approx}
\bar{\bm{R}}_\textrm{s} \approx \bar{\bm{R}}_\textrm{t} - \sigma^2 \bm{I}_M,
\end{equation}
where $\sigma^2$ denotes the noise variance. On one hand, the approximation in (\ref{R_approx}) is due to the cross-correlation between channel coefficients. On the other hand, the insufficient time and frequency samples may also affect the approximation since (\ref{R_approx}) is met exactly based on ensemble-average but not so with time-average CCM. Note that we assume the average uplink transmit power of all users is identical.  The spatial compression efficiency \cite{ama15} \cite{sae15}, which is defined as the ratio between the reserved signal power by a limited number of RF chains after spatial compression including beamspace transformation and proposed beam combination, and the total receive signal power in a given signal block; it is written as
\begin{equation}
\label{eta}
\eta\left(\bm{A}_\mathcal{C}\right) = \frac{\textrm{tr}\left[\bm{A}_\mathcal{C} \bm{A}_\mathcal{L} \bar{\bm{R}}_\textrm{s} \bm{A}_\mathcal{L}^\dag \bm{A}_\mathcal{C}^\dag\right]}{\textrm{tr} \bar{\bm{R}}_\textrm{s} },
\end{equation}
where it is prescribed that $\bm{A}_\mathcal{C}$ has orthonormal rows such that the transform efficiency is well defined with range $\eta \in [0,1]$. The signal after spatial compression is $\bm{y}^\prime \triangleq \bm{A}_\mathcal{C}\bm{A}_\mathcal{L}\bm{y}$, and therefore the numerator of \eqref{eta} is the reserved signal power after spatial compression. Since the beamspace transformation $\bm{A}_\mathcal{L}$ is considered to be fixed in this paper, the spatial compression efficiency is only a function of beam combination matrix $\bm{A}_\mathcal{C}$.

\subsection{Spatial Compression Procedure}
The central goal of the paper is to design $\bm{A}_\mathcal{C}$ subject to the hardware constraints in \eqref{ac_const}, so as to further reduce the number of RF chains, i.e., $K$. Towards this end, the general procedure to obtain the combination weights is described below. The detailed algorithm is illustrated later in Section \ref{sec:st}.
\subsubsection{Channel Estimation}
First, the CCM after beamspace transformation is obtained by sweeping over all beams. This is achieved by setting the phase shifters to off state or zero phase (corresponding to unity) to realize beam switching. Advanced algorithms which leverage the compressive-sensing technique and avoid a complete beam sweeping can be found in, e.g., \cite{gao17_ce}.
\subsubsection{Beam Combination Weights Determination}
After the CCM is obtained, the combination weights $\bm{A}_\mathcal{C}$ is determined by the proposed algorithms described in Section \ref{sec:st}.
\subsubsection{Data Transmission}
Then, the baseband digital processing is performed over the beam-domain signal after beam combination. The time duration between adjacent beam combination weights design is related to the CCM variation speed ($1$~second to $10$~seconds \cite{Adhikary13}), which is in general much slower than instantaneous CSI (micro-seconds). Therefore, the channel estimation overhead is relatively low.

\section{Motivating Examples: Why Beamspace Transformation is not Enough for RF Chain Reduction?}
\label{sec:me}
The beamspace transformation takes advantages of the angular power sparsity of massive MIMO channels to reduce the number of RF chains by selecting the most significant signal directions \cite{brady13}. The theoretical support of the approach stems from the fact that the optimal spatial compression scheme is proved to be SVD-based \cite{sae15}, and that the DFT-based beamspace transformation is asymptotically equivalent to SVD approach when the number of antennas is large and the time averaged CCM estimation equals the ensemble-averaged CCM \cite{Adhikary13}. In what follows, examples are given to demonstrate that when these two conditions are not met exactly in practice, we can further reduce the number of RF chains by combining correlated beams which essentially  experience correlated propagation channels.

\begin{figure}[!t]
	\centering
	\includegraphics[width=0.45\textwidth]{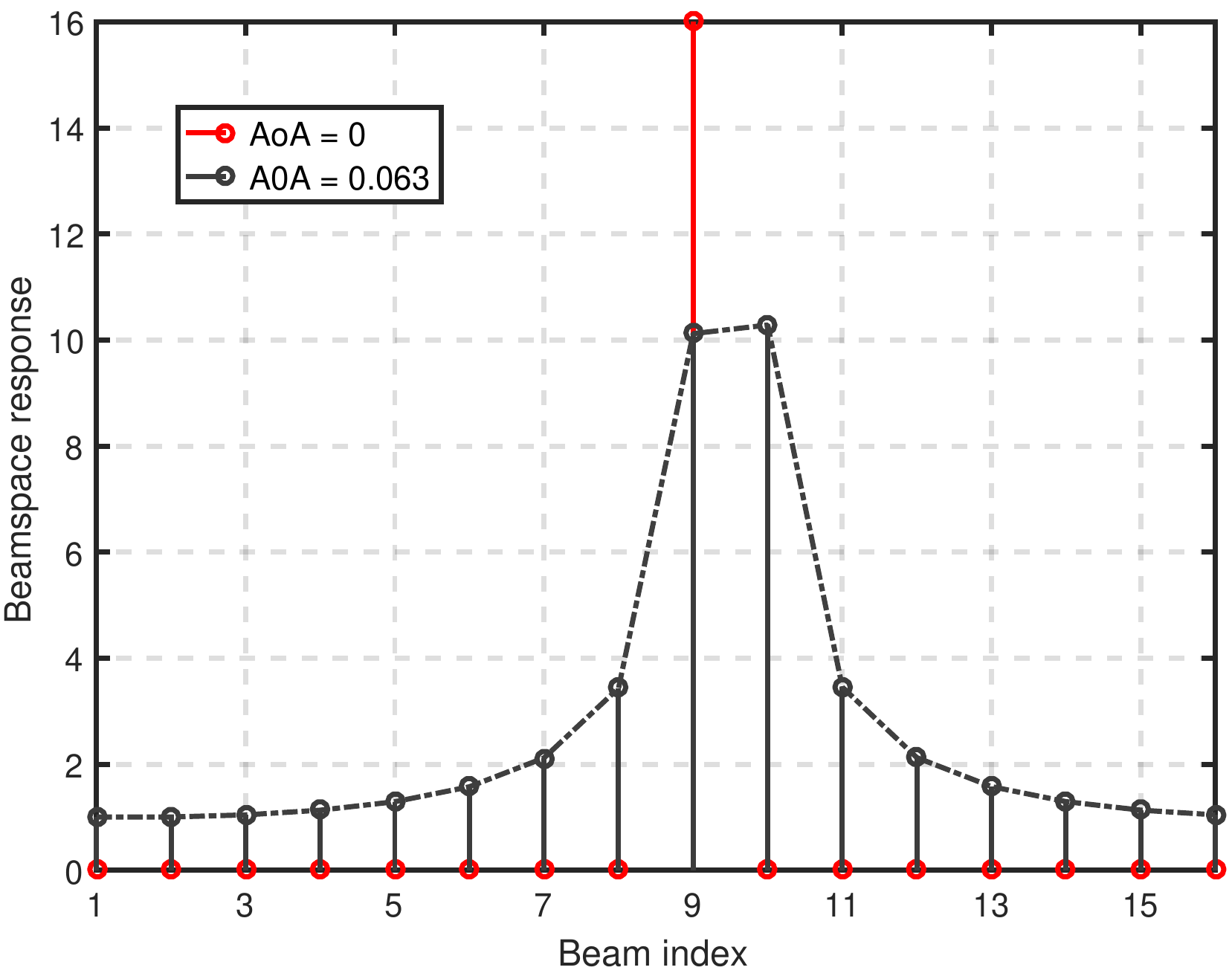}
	\caption{An example with $16$ antennas case. The channel beamspace response is shown with only one MPC. The AoAs are different to demonstrate the DFT power leakage issue.}
	\label{Fig_dft_leak}
\end{figure}

It is a known fact that the DFT suffers from power leakage, especially when the number of the DFT points is limited. In \cite{gao16_bs}\cite{zeng16}, it is proved that the power leakage $P_\textrm{leak}$ of a beam given the AoA $\theta$ and beam index is approximated by
\begin{equation}
P_\textrm{leak} \sim \textrm{sinc}^2 \left(m-\frac{d\sin(\theta)}{\lambda}\right),
\end{equation}
where $\textrm{sinc}(x) \triangleq \frac{\sin(x)}{x}$ and $m$ is the beam index. Fig. \ref{Fig_dft_leak} shows that when the AoA coincides with the AoA of some DFT vector, e.g., AoA is zero corresponding to the first DFT vector, then only one beam after the beamspace transformation can perfectly contain all the signal power. In this case, only one RF chain is required with spatial compression efficiency of $1$. However, the case with a slightly different AoA wherein $\theta = 0.063$ shows a significant power leakage. In this circumstance, more than one RF chains are needed to achieve a target compression efficiency. Previous work \cite{gao16_bs}\cite{zeng16} usually assumes the ideal case which ignores this effect by assuming the AoAs are always matched to the DFT vectors. Based on this example, it can be anticipated that there is potential to reduce the number of RF chains beyond beamspace transformation in general propagation environments.

On the other hand, it is likely that the time-averaged CCM in \eqref{ccm_cal} \eqref{ccm_u} does not converge to the ensemble-average CCM in \eqref{en_CCM} due to finite time and frequency samples which could also lead to potential RF chain waste. Concretely, suppose a channel vector with $3$ MPCs in total. If there is infinite samples, in theory the effective rank (effectively significant rank) of the CCM should be $3$ assuming i.i.d. fading for each MPC. However, consider an extreme example where the channel is static in the given signal block, which happens when the user and the scatterers are both static during the time, and consequently the rank of the CCM estimated in the signal block is $1$. As a result, one beam is sufficient. In the mean time, there are still $3$ different MPCs with distinct AoAs, and hence the beamspace transformation detects $3$ beams (ignoring spatial leakage). In this case, $2$ RF chains are wasted and they could be saved for less RF complexity. 

Based on the insights provided in the above examples, we propose to adopt a beam combination module after the lens antenna array to further reduce the number of required RF chains. Furthermore, hardware constrains are considered where limited-resolution digital phase shifters are adopted.

\section{Beam Combination Schemes}
\label{sec:st}
\subsection{Spatial Compression without Hardware Constraints}
The following theorem addresses the question: \textbf{What are the optimal beam combination weights to maximize the spatial compression efficiency without considering hardware constraints?} The answer serves as a performance benchmark for schemes with hardware constraints.
\begin{theorem}
	\label{thm3}
	With finite samples of receive signal vectors of dimension $N$, i.e.,  $\bm{y}_{l,t}$, $(l=1,...,L$, $t=1,...,T)$, denote the first $N_\textrm{s}$ ($N_\textrm{s} \le N$) columns of the singular matrix of $\bar{\bm{R}} =\frac{1}{TL}\sum_{t,\,l} \bm{y}_{l,t}\bm{y}_{l,t}^\dag$ as $\bm{F}_{\textrm{U}}$, and the set $\mathcal{F} = \left\{\bm{F}:\,\bm{F} = \bm{V}\bm{F}_{\textrm{U}}^\dag,\,\bm{V}\bm{V}^\dag = \bm{V}^\dag\bm{V}=\bm{I}_{N_\textrm{s}}\right\}$. For any $\bm{F}_{\textrm{opt}} \in \mathcal{F}$,
	\begin{equation}
	\bm{F}_{\textrm{opt}} = \argmax_{\bm{F} \in \mathbb{H}_{N_\textrm{s}}}\,\eta(\bm{F}),
	\end{equation}
	and,
	\begin{equation}
	\eta(\bm{F}_{\textrm{opt}}) = \frac{\sum_i^{N_\textrm{s}}\left({\lambda_i}-\sigma^2 \right) }{\sum_i^{N}\left({\lambda_i} - \sigma^2 \right)},
	\end{equation}
	where $\mathbb{H}_{N_\textrm{s}}$ denotes the space of all $N_\textrm{s} \times N $ matrices $\bm{F}$ with orthonormal rows, $\lambda_i\,\left(i=1,...,N\right)$ are the singular values of $\bar{\bm{R}}$, $\sigma^2$ is the noise variance, and $\eta(\bm{F})$ is defined in \eqref{eta}.
\end{theorem}

\begin{IEEEproof}
	See Appendix \ref{app2}.
\end{IEEEproof}

\subsection{Spatial Compression with Hardware Constraints}
Based on Theorem \ref{thm3}, the optimum beam combination is the first $K$ dominant eigenvectors of the time-averaged CCM (or left multiplied by a unitary matrix). To perform this optimal beam combination, a total of $M$, which equals the number of antennas, infinite-resolution phase shifters with variable amplitude are required since the weights are arbitrary. Therefore, it entails high cost and complexity. To reduce the cost of the beam combination module, two ideas are exploited. First, since the beamspace channel is sparse, conventional beam selection realized by a switching network as in Fig. \ref{Fig_arch} can be adopted \cite{7}. Second, an additional limited-resolution PSN with constant amplitude (assumed to be unity in the paper) is added to combine correlated beams to further compress the beamspace channel dimensionality. To make the first idea concrete, the central problem is that \textbf{how many beamspace beams are needed}. In this regard, the following proposition is presented. 

\begin{prop}
	\label{thm2}
	Consider one-dimensional lens antenna array with $M$ critically placed antennas. In the large array regime, i.e., $M \to \infty$, the number of non-zero beams after beamspace transformation is 
	\begin{equation}
	\mathcal{D}_\textrm{T}  \stackrel{M \to \infty}{\longrightarrow} \frac{M}{2} \bigcup_n \Omega_n + \smallO(M) ,
	\end{equation}
	where $\Omega_n$ is defined as the signal angular spread of the $n$-th user in terms of directional sines.
\end{prop}
\begin{IEEEproof}
	See Appendix \ref{pr_th1}.
\end{IEEEproof}

\begin{remark}
	The scaling result of the number of non-zero power beams after beamspace transformation with the number of antennas is given in Proposition \ref{thm2}. It can be leveraged to determine the number of retained beams after the beam selection. Concretely, suppose there are two users with angular spread of $[-\frac{\pi}{3},-\frac{\pi}{6}]$ and $[\frac{\pi}{4},0]$, respectively. Then 
	\begin{IEEEeqnarray}{rCl}
	\mathcal{D}_\textrm{T}  &\stackrel{M \to \infty}{\longrightarrow}& \frac{M}{2} \bigcup \left\{\left[\sin\left(-\frac{\pi}{3}\right),\sin\left(-\frac{\pi}{6}\right)\right], \left[\sin\frac{\pi}{4},0\right]\right\} \nonumber\\
	&=& \frac{\sqrt{3}}{4}M.
	\end{IEEEeqnarray}
\end{remark}

After the beam selection by the switching network, the selected beams are combined to further reduce the number of RF chains by a finite-resolution PSN. The problem of maximizing the spatial compression efficiency subject to \textbf{hardware constraint} is formulated by
\begin{flalign}
\label{p2}
\textbf{P1:}&&\mathop{\textrm{maximize}}\limits_{\bm{A}_\mathcal{C}}  \,\,& \eta\left(\bm{A}_\mathcal{C}\right) = \frac{\textrm{tr}\left[\bm{A}_\mathcal{C} \bm{A}_\mathcal{L} \bar{\bm{R}}_\textrm{s} \bm{A}_\mathcal{L}^\dag \bm{A}_\mathcal{C}^\dag\right]}{\textrm{tr} \bar{\bm{R}}_\textrm{s} } &&\\
\label{p2_const}
&&\textrm{s.t.,}\,\, &  \left[\bm{A}_\mathcal{C}\right]_{i,j} \in \bm{\Psi}, \nonumber \\
&&& \bm{\Psi} = \left\{e^{\frac{j2n\pi}{2^B}},\,n = 0,...,2^B-1\right\}.&&
\end{flalign}
It is observed that the problem is a combinatorial problem with a large scale, which is generally NP-hard. In this paper, we adopt a BB-based approach to solve for the optimum solution which, admittedly, has a high complexity but with optimality. Therefore, it can be viewed as a performance upper bound for the other low-complexity heuristic algorithms. The BB algorithm is a widely-used method to solve discrete programming problem which is guaranteed to converge to optimum \cite{land60}. However, the most critical challenge in developing a BB-based approach is to \emph{solve the sub-problem} in order to find an appropriate bound for each branch, hence the name ``branch and bound''. Interested readers can see, e.g., \cite{israel13}\cite{land60}, for the details about the BB algorithm. Without further complications, the sub-problem is a problem about what are the optimum weights when a subset of the weights is given, which can be formulated as
\begin{flalign}
\label{p1}
\textbf{P2:}&&\mathop{\textrm{maximize}}\limits_{\bm{w}_\textrm{J}}  \,\,& \eta(\bm{x}) \triangleq \frac{\bm{x}^\dag \bm{R} \bm{x}}{\bm{x}^\dag \bm{x}} &&\\
\label{p_const}
&&\textrm{s.t.,}\,\,  & \bm{x} = 
\left[ {\begin{array}{*{20}{c}}
	\bm{d}_\textrm{I}\\
	\bm{w}_\textrm{J}
	\end{array}} \right],&&
\end{flalign}
where $\bm{R}$ is Hermitian positive semi-definite, $\bm{x}\in\mathbb{C}^{L}$, $\bm{d}_\textrm{I}\in\mathbb{C}^{l}$ is a given complexed-valued vector, and $1 \le l \le L$. For ease of exposition, denote 
\begin{equation}
\bm{R} = \left[ {\begin{array}{*{20}{c}}
	{{\bm{R}_\textrm{I}}}&{{\bm{R}_\textrm{IJ}}}\\
	{{\bm{R}_\textrm{JI}}}&{{\bm{R}_\textrm{J}}}
	\end{array}} \right],
\end{equation}
where ${\bm{R}_\textrm{IJ}} = \bm{R}_\textrm{JI}^\dag$. The SVD of $\bm{R}_\textrm{J}$ is $\bm{R}_\textrm{J} = \bm{U}_\textrm{J} \bm{\Sigma}_\textrm{J} \bm{U}_\textrm{J}^\dag$, where $\bm{\Sigma}_\textrm{J} = \textrm{diag}[\lambda_1,...,\lambda_{L-l}]$ and $\lambda_1 \ge \lambda_2 \ge ... \ge \lambda_{L-l}$. Denote $\bm{u}_\textrm{J,dom}$ as one of the dominant singular vector of $\bm{R}_\textrm{J}$, $\bm{p} \triangleq \bm{R}_\textrm{JI}\bm{d}_\textrm{I}$, $r \triangleq \bm{d}_\textrm{I}^\dag \bm{R}_\textrm{I} \bm{d}_\textrm{I}$, and $d \triangleq \bm{d}_\textrm{I}^\dag \bm{d}_\textrm{I}$.

\begin{figure}[!t]
	\centering
	\includegraphics[width=0.45\textwidth]{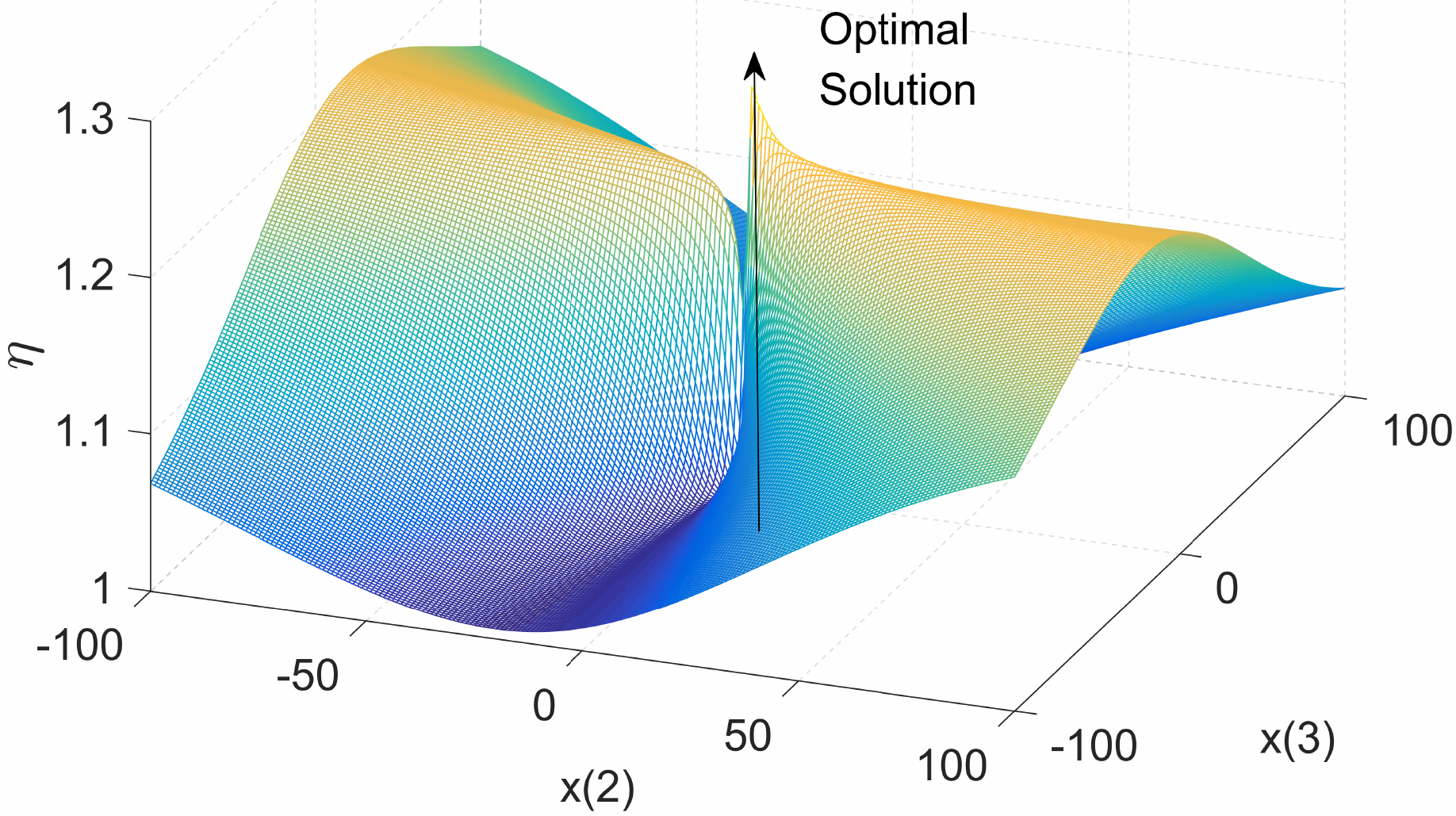}
	\caption{An example of the sub-problem objective function with $\bm{R}$ is a randomly generated three-dimensional CCM. $\bm{d}_\textrm{I} = 1$ and $x(2)$, $x(3)$ are x- and y-axis, respectively.}
	\label{Fig_subp}
\end{figure}

This sub-problem is directly derived from solving \textbf{P1} step by step by the BB method, and relax the discrete constraints to continuous to obtain an upper bound. Concretely, the optimization in \textbf{P2} is over $\bm{x}$ which is one column of $\bm{A}_\mathcal{C}$, given $\bm{R}$ as the CCM after beamspace transformation and beam selection, i.e., $\bm{R} = \bm{A}_\mathcal{L} \bar{\bm{R}}_\textrm{s} \bm{A}_\mathcal{L}^\dag$. We emphasize that even without the discrete constrains, the sub-problem \textbf{P2}, being a non-convex problem since the objective function is not concave, is still very difficult to solve. An example of the sub-problem objective function is depicted in Fig. \ref{Fig_subp}, where it is observed that the global optimum solution is not attainable by a commonly-used, e.g., gradient-ascend-based method. In the following theorem, we derive the \textbf{optimum solution} in a closed-form (given the SVD of $\bm{R}$) based on a constructive proof, wherein we hypothesis the solution has a special structure and prove that such a structure is indeed the optimum solution.

\begin{theorem}
	\label{thm_opt}
	The optimum objective value of \textbf{P2} is given by
	\begin{IEEEeqnarray}{rCl}
	\eta^*(\bm{d}_\textrm{I}) &\triangleq&  \limsup_{\bm{w}_\textrm{J} \to \bm{w}_\textrm{J}^*} \eta(\bm{x}) \nonumber\\
	&=& \left\{ {\begin{array}{*{20}{l}}
		{\max[\lambda_1,r/d],\;{\textrm{if}}\; \bm{u}_\textrm{J,dom}^\dagger \bm{p} = 0\textrm{ and C1}}\\
		{\lambda^*,\;{\textrm{otherwise}},}
		\end{array}} \right. 
	\end{IEEEeqnarray}
	where the condition C1 is
	\begin{flalign}
    \label{condition1}
    \textrm{C1}:&&\lambda_1 d - r - \sum_{i=m+1}^{L-l} \frac{\left|\left(\bm{U}_\textrm{J}^\dagger \bm{p}\right)_i\right|^2}{\lambda_1-\lambda_i} > 0,&&
    \end{flalign}
	and $m$ is the dimensionality of the dominant singular subspace of $\bm{R}_\textrm{J}$, and $\lambda^*$ satisfies
	\begin{equation}
	\lambda_1 < \lambda^* \le \frac{d\lambda_1+r+\sqrt{(d\lambda_1-r)^2+4d\bm{p}^\dagger\bm{p}}}{2d},
	\end{equation} 
	and $\lambda^*$ is the unique solution of the equation
	\begin{equation}
	\sum_{i=1}^{L-l} \frac{\left(\bm{U}_\textrm{J}^\dagger \bm{p}\right)_i}{\lambda-\lambda_i} = \lambda d - r.
	\end{equation}
	The optimum solution $\bm{w}_\textrm{J}^*$ is 
	\begin{equation}
	\bm{w}_\textrm{J}^* = \left\{\,
	    \begin{IEEEeqnarraybox}[][c]{l?s}
	    \IEEEstrut	
		\beta \bm{u}_\textrm{J,dom}\textrm{ or }\bm{0}, & if $\bm{u}_\textrm{J,dom}^\dagger \bm{p} = 0$ and C1,\\
		\left(\lambda^* \bm{I}_{L-l} - \bm{R}_\textrm{J}\right)^{-1}\bm{p}, & otherwise,
		\IEEEstrut
		\end{IEEEeqnarraybox}
		\right.
	\end{equation}
	where $\beta \to \infty$.
\end{theorem}
\begin{IEEEproof}
See Appendix \ref{app1}.
\end{IEEEproof}

\begin{remark}
    It is noteworthy that the limiting case of Theorem \ref{thm_opt}, i.e., $\bm{u}_\textrm{J,dom}^\dagger \bm{p} = 0$ and C1, almost never happens in practice since the condition is very strict. Therefore, the case is derived more for mathematical completeness rather than practical concerns.
\end{remark}
\begin{corollary}
	\label{coro1}
	An approximation of the optimum solution of \textbf{P2} with discrete constraints in \eqref{ac_const} is 
	\begin{IEEEeqnarray}{rCl}
	&& \eta_\textrm{appox}(\bm{d}_\textrm{I}) \nonumber\\
	&=& \frac{\bm{d}_\textrm{I}^\dag \bm{R}_\textrm{I} \bm{d}_\textrm{I}+ (\bm{w}_\textrm{J}^*)^\dag \bm{R}_\textrm{J} {\bm{w}}_\textrm{J}^* + \bm{p}^\dag {\bm{w}}_\textrm{J}^* + ({\bm{w}}_\textrm{J}^*)^\dag \bm{p}+ \sigma_{\bm{e}}^2 \frac{\textrm{tr}\bm{R}_\textrm{J}}{L-l} }{\bm{d}_\textrm{I}^\dag \bm{d}_\textrm{I}+({\bm{w}}_\textrm{J}^*)^\dag {\bm{w}}_\textrm{J}^* +  \sigma_{\bm{e}}^2}, \nonumber\\
	\end{IEEEeqnarray}
where
\begin{IEEEeqnarray}{rCl}
	\sigma_{\bm{e}}^2 &=&  2^{-B} (\bm{w}_\textrm{J}^*)^\dag \bm{w}_\textrm{J}^*.
\end{IEEEeqnarray}
\end{corollary}
\begin{IEEEproof}
	The proof is based on the rate-distortion theory. See Appendix \ref{app3} for details.
\end{IEEEproof}

Based on Theorem \ref{thm_opt} and Corollary \ref{coro1}, we are ready to develop the BB-based discrete beam combination (BB-BC) scheme. It is described in Algorithm \ref{alg:bbb}.
\begin{algorithm}[!t]
	\caption{Branch and Bound Based Beam Combination (BB-BC)}
	\label{alg:bbb}
		\KwIn{Channel correlation matrix estimation $\bm{R}$;
		The number of combined beams $K$;}
		\KwOut{The beam combination matrix, $\bm{A}_\mathcal{C}$;}
		\label{alg:init}
		Initialization: Set $\bm{R}_\textrm{b} = \bm{R}$, $\bm{A}_\mathcal{C} = \phi$.\\
		\label{alg:branch}
		\For{b=1:K}{
			Set $\mathcal{G} = \left\{\left(\bm{\Psi}^L,\phi\right)\right\}$. Set $\hat{\eta} = f(\bar{\bm{u}}_1,\bm{R})$, where $\bar{\bm{u}}_1$ is the dominant eigenvector of $\bm{R}_\textrm{b}$ rounded to the nearest element in $\bm{\Psi}$ (entry-wise). Set $\bm{w}_b = \bar{\bm{u}}_1$. \\
			\For{$\mathcal{G} \neq \phi$}{
				Branch: Choose $\bm{S} \in \mathcal{G}$, which satisfies $\bm{S} = \argmax_{\bm{X}\in\mathcal{G}} \eta(\bm{X}_2)$. 
				Partition $\bm{S}$ into $\bm{S}_1$,..., $\bm{S}_{2^B}$, where $\bm{S}_i$ is the set satisfying $(\bm{S}_i)_2 = [(\bm{S})_2^T,\psi_i]^T$. \\
				Bound: Set $\mathcal{G} \leftarrow \mathcal{G} \backslash \bm{S}$. \For{$1 \le i \le 2^B$}{Calculate $\eta_i = f((\bm{S}_i)_2,\bm{R}_\textrm{b})$, and $\bm{w^\prime}_i$ is the corresponding optimum solution. \\
					\If{$\eta_i > \hat{\eta}$}{
						\If{$\mathcal{L}((\bm{S}_i)_2)=L$}{$\hat{\eta} = \eta_i$, $\bm{w}_b = (\bm{S}_i)_2$.}
						\Else{Set $\mathcal{G} \leftarrow \mathcal{G} \cup \bm{S}_i$\\
						\label{alg:retain}
						Round $\bm{w}^\prime_i$ to the neaest point in $\bm{\Psi}$ as $\bar{\bm{w}}^\prime_i$. \\
						\If{$\eta(\bar{\bm{w}}^\prime_i) > \hat{\eta}$}{$\hat{\eta} = \eta(\bar{\bm{w}}^\prime_i)$, $\bm{w}_b = \bar{\bm{w}}^\prime_i$}}
						}	
				} 				
	    	}
    		Set $\bm{A}_\mathcal{C} \leftarrow [\bm{A}_\mathcal{C},\bm{w}_b]$, \\
    		$\bm{R}_\textrm{b} \leftarrow \left(\bm{I}_L - \frac{1}{L}\bm{A}_\mathcal{C}\bm{A}_\mathcal{C}^H\right)^\dag \bm{R}_\textrm{b} \left(\bm{I}_L - \frac{1}{L}\bm{A}_\mathcal{C}\bm{A}_\mathcal{C}^H\right)$.
    		\label{R_proj}    		
   		}	
		\Return $\bm{A}_\mathcal{C}$.
\end{algorithm}
For $\bm{X}\in\mathcal{G} $, $\bm{X}_1$ and $\bm{X}_2$ denote the first and second entries of $\bm{X}$, respectively. $\psi_i$ is the $i$-th entry of $\bm{\Psi}$. $f(\bm{d},\bm{R})$ is defined as the optimum objective value of \textbf{P2} with CCM $\bm{R}$ and $\bm{d}_\textrm{I} = \bm{d}$. The length of a vector $\bm{x}$ is denoted by $\mathcal{L}(\bm{x})$. $\eta(\bm{x})$ is defined in \textbf{P2}. The number of beams after beam selection is denoted by $L$. 

The initial feasible solution is obtained by rounding the SVD-based solution to the nearest point in $\bm{\Psi}$, and the initial lower bound of the optimum is thus the objective function evaluated at this rounded point. The BB-BC works roughly as follows. We design beam combination weights for each column of the matrix $\bm{A}_\mathcal{C}$ successively and project to the orthogonal subspace of the CCM after selecting one column to avoid repetitive selection as in the \ref{R_proj}-th step. It can be easily verified that 
\begin{equation}
\left(\bm{I}_L - \frac{1}{L}\bm{A}_\mathcal{C}\bm{A}_\mathcal{C}^H\right) \bm{x} = 0,\, \forall \bm{x} \in \textrm{range}(\bm{A}_\mathcal{C}),
\end{equation}
where $\textrm{range}(\bm{A}_\mathcal{C})$ denotes the column space of $\bm{A}_\mathcal{C}$. In each step, the BB-based scheme first branches on the existing candidate sets, each of which is possible to contain the optimum solution. The branch criterion is to select one that is the mostly likely, based on the optimum objective function value $f(S_2,\bm{R})$ of each set $\bm{S}$ by Theorem \ref{thm_opt}. Compared with other branch criterion, e.g., width-first-search (branch the set with the smallest number of determined weights) or depth-first-search (branch the set with the largest number of determined weights), the adopted best-first approach shows better performance in terms of faster convergence in our simulations. After the branching, the branched sets are compared with the current best feasible solution by solving the continuous sub-problem for each set. The idea is that if the upper bound of the set is not as good as the current best feasible solution, it is unnecessary to keep branching that set. Therefore, the set is eliminated. Only the ones whose upper bound is better than the current best are retained as in the \ref{alg:retain}-th step. Meanwhile, we update the current best by rounding the optimum solution of the sub-problem if it is better. The algorithm continues until there is no more set to be branched. 

It can be observed that the BB-BC searches over all the possible sets, and therefore is guaranteed to find the optimum solution. To accelerate the algorithm, one can adopt an alternative stopping criterion which ensures that the obtained maximum is near the optimum. The criterion can be written as
\begin{equation}
\label{tc}
f(\bm{S}_2,\bm{R}_\textrm{b}) - \hat{\eta} < \epsilon \hat{\eta},\,\forall \bm{S} \in \mathcal{G}.
\end{equation}
Then 
\begin{equation}
\hat{\eta} > \frac{1}{1+\epsilon} \max_{\bm{S} \in \mathcal{G}} f(\bm{S}_2,\bm{R}_\textrm{b}) >\frac{1}{1+\epsilon} \eta^*.
\end{equation}
Hence the termination criterion in \eqref{tc} ensures the resultant feasible solution is within $1/(1+\epsilon)$ of the optimum. Moreover, the Corollary \ref{coro1} can be used to obtain an approximation of the upper bound $\hat{\eta}$ to further accelerate the convergence.

\begin{figure}[!t]
	\centering
	\includegraphics[width=0.45\textwidth]{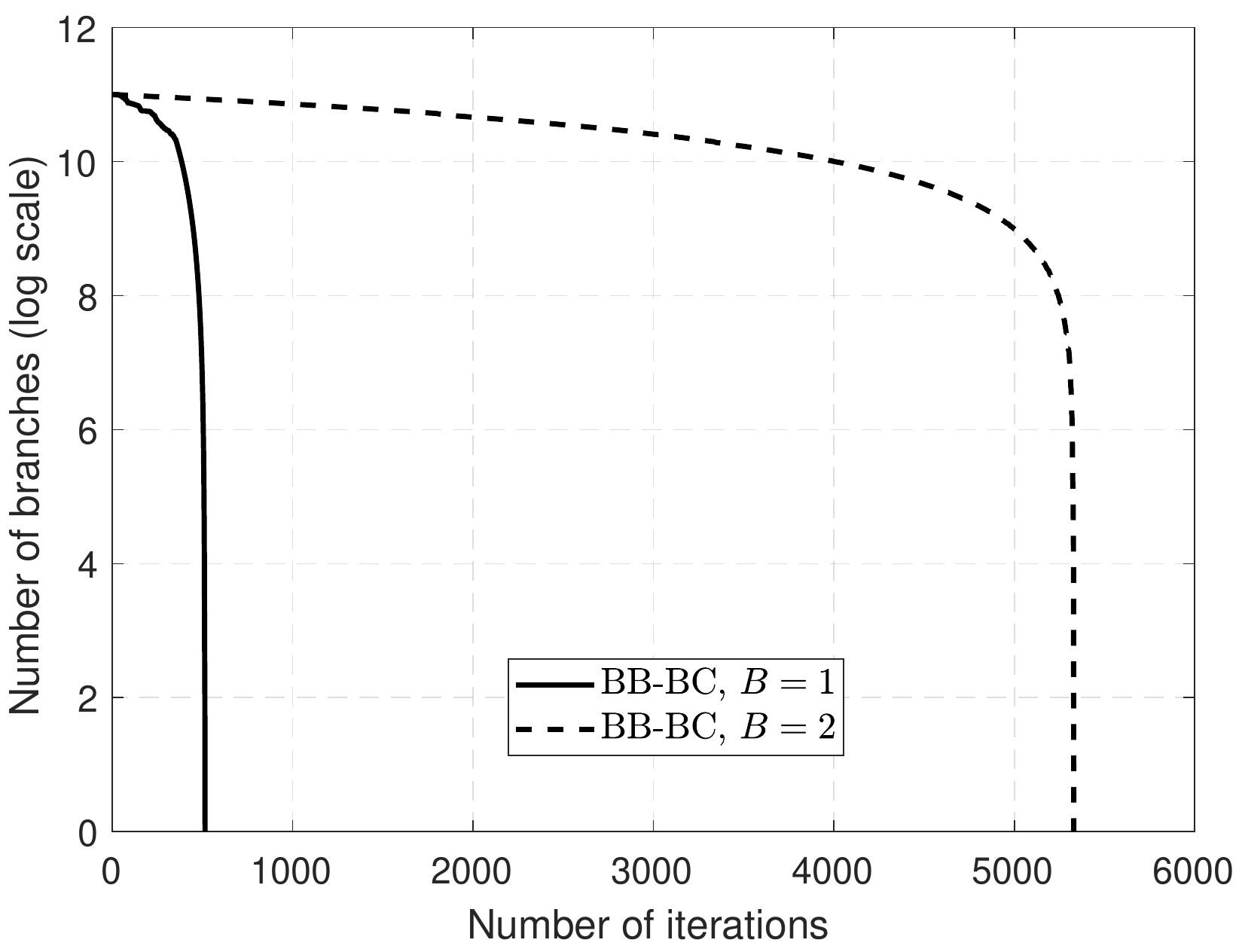}
	\caption{The number of remaining branches when running BB-BC (determining one column) with $L=12$. The number of users is $2$, and the user moving speed is $3$~km/h. UL SNR is $20$~dB.}
	\label{Fig_branch}
\end{figure}

The computational complexity of the BB-BC scheme is determined by two factors. The first is the number of remaining branches after each iteration of BB-BC. This is illustrated in Fig. \ref{Fig_branch}. The number of remaining branches ($y$) is denoted in the logarithm scale, i.e., $\log_{2^B}(y)$,  as the $y$-axis. The total number of branches for exhaustive search is $2^{B(L-1)}$, by noticing that the spatial compression efficiency is insensitive to a constant phase rotation. It can be observed that the number of required iterations to find the optimum is significantly smaller than the total number of branches by exhaustive search (about $5000$ compared with $2^{22}$ when $B=2$), thanks to the branch-and-bound operations. Secondly, the computational complexity in each iteration can be upper bounded by $2^B L^3$ assuming each bounding operation is performed on an $L$-dimensional CCM.

Even though the BB-BC method alleviates the computational complexity by dynamically eliminating the unqualified branches, it is still very time-consuming and computation demanding, especially when the number of antenna elements is large. Towards this end, the SG-BC scheme is proposed, which is essentially a heuristic method which selects the beam combination weights sequentially based on Theorem \ref{thm_opt}. Therefore, the complexity scales linearly with the number of beams, compared with exponentially for the BB-BC scheme. In Algorithm \ref{alg:sgs}, the SG-BC is described.

\begin{algorithm}[!ht]
	\caption{Sequential Greedy Beam Combination (SG-BC)}
	\label{alg:sgs}
	\KwIn{Channel correlation matrix estimation $\bm{R}$;
		The number of combined beams $K$;}
	\KwOut{The beam combination matrix, $W$;}
	Initialization: Set $\bm{R}_\textrm{b} = \bm{R}$, $\bm{A}_\mathcal{C} = \phi$.\\
	\For{b=1:K}{
		Set $\bm{w}_b=\phi$.\\
		\For{l=1:L}{
			$w_l=\argmax_{\phi_i \in \bm{\Psi}} \eta([\bm{w}_b^T, \phi_i]^T)$.\\
			$\bm{w}_b \leftarrow [\bm{w}^T,w_l]^T$
		}
		Set $\bm{A}_\mathcal{C} \leftarrow [\bm{A}_\mathcal{C},\bm{w}_b]$, \\
		$\bm{R}_\textrm{b} \leftarrow \left(\bm{I}_L - \frac{1}{L}\bm{A}_\mathcal{C}\bm{A}_\mathcal{C}^H\right)^\dag \bm{R}_\textrm{b} \left(\bm{I}_L - \frac{1}{L}\bm{A}_\mathcal{C}\bm{A}_\mathcal{C}^H\right)$.
	}	
	\Return $\bm{A}_\mathcal{C}$.
\end{algorithm}

\begin{remark}
	The SG-BC scheme can be viewed as a best-\emph{only} search BB-based algorithm. Instead of searching over all the branches, it only selects the best branch and discards the rest, by solving the sub-problem based on Theorem \ref{thm_opt}. 
\end{remark}
\section{Simulation Results}
\label{sec:sr}
In this section, to test our proposed compression schemes, we will present simulation results using a link-level simulator based on the LTE numerology and 3GPP spatial channel models (SCMs) \cite{baum05}. The parameters are specified in Table \ref{table_para}. The spatial compression efficiency in \eqref{eta} is adopted to evaluate the performance. Note that based on \eqref{eta}, we do not distinguish between useful signal and interference but focus purely on the retained signal power after spatial compression, due to the fact that the proposed spatial compression module is implemented in the RF and hence assumed to have no knowledge of the interference statistics. The spatial compression module takes time-domain signals as input and outputs the compressed dimensionality-reduced signal streams. The subsequent signal processing modules, such as orthogonal-frequqncy-division-multiplexing (OFDM) demodulation, decoding and etc., are exactly the same as the conventional LTE systems.
\begin{table}[!t]
	\renewcommand{\arraystretch}{1.3}
	\caption{Simulation Parameters}
	\label{table_para}
	\centering
	\begin{tabular}{| l | r|}
		\hline
		Carrier frequency & $2.6$~{GHz}  \\
		\hline
		System bandwidth & $20$~{MHz}  \\
		\hline
		Subcarrier spacing & $15$~{KHz}  \\
		\hline
		OFDM FFT size & $2048$ \\
		\hline
		BS antenna spacing & $0.5\lambda$ \\
		\hline
		User antennas & $1$ omni-directional \\
		\hline
		User mean DoAs & uniformly distributed \\
		\hline
		Channel model & 3GPP Urban Micro \\		
		\hline
		Number of rays & $6$ \\
		\hline
		Angular spread of each ray  &  $5$ degrees \\
		\hline
		Total angular spread for all rays  & $45$ degrees\\
		\hline
		Delay spread & $700$~{ns} \\
		\hline
		Simulation time & $400$~{ms}\\
		\hline
	\end{tabular}
\end{table}

\begin{figure}[!t]
	\centering
	\includegraphics[width=0.45\textwidth]{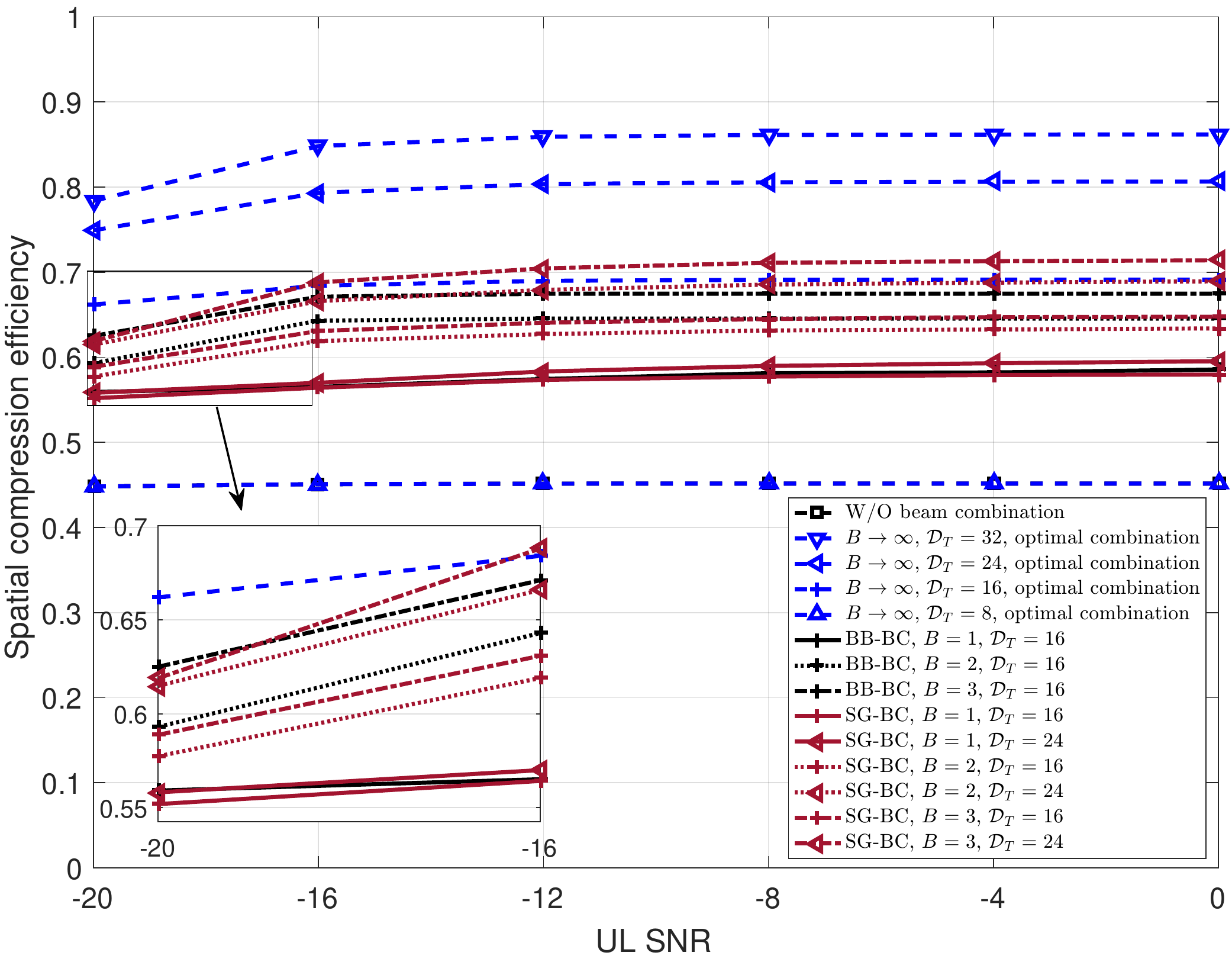}
	\caption{Comparisons of proposed BB-BC, SG-BC, beamspace transformation without beam combination and optimal combination schemes with $128$ antenna ports and $8$ RF chains. The number of users is $2$, and the user moving speed is $3$~km/h.}
	\label{Fig_pr_A128_B8_Bd16_v3_n2}
\end{figure}

In Fig. \ref{Fig_pr_A128_B8_Bd16_v3_n2}, the comparison is made among the proposed schemes BB-BC and SG-BC, and the optimal combination scheme given by Theorem \ref{thm3} and beamspace transformation without beam combination. The baseline, i.e., performance without beam combination, is obtained by selecting a number of the strongest beams without beam combination. First, it is observed that beam combination after beamspace transformation is able to improve the spatial compression efficiency with the same number of RF chains. It is mainly due to spatial power leakage and imperfect channel statistics estimation which have been explained in Section \ref{sec:me}. Even with stringent hardware constraints, i.e., the resolution of digital phase shifters is limited and the amplitude is constant, the performance improvements over the one without beam combination is obvious, enabling us to adopt the proposed low-resolution PSN. It is seen that a one-bit PSN already improves the spatial compression efficiency by about $10\%$, and that a $2$-bit PSN improves by $20\%$. Furthermore, a $3$-bit PSN only has marginal performance advantage over $2$-bit, meaning that a ``dirty'' low-resolution PSN is sufficient. On the other hand, the UL signal-to-noise-ratio (SNR) has little impact on the compression efficiency performance because the uplink CCM is estimated with a large number of OFDM symbols, which are from one LTE subframe and the whole bandwidth, i.e., $14 \times 1200 = 16800$ symbols. Note that the UL SNR is the per-antenna received SNR, and therefore the SNR of each beam after the beamspace transformation is much larger, e.g., $M$ antenna elements bring about $10\log_{10}M$~dB beamforming gain \cite{tse05}. 

The other important note is that the performance of the SG-BC scheme is close to the optimal BB-BC scheme with hardware constraints. Given the dramatic complexity reduction by the SG-BC scheme (linear with the number of beams after beamspace transformation compared with exponential). It is much more desirable in practice. 
\begin{figure}[!t]
	\centering
	\includegraphics[width=0.45\textwidth]{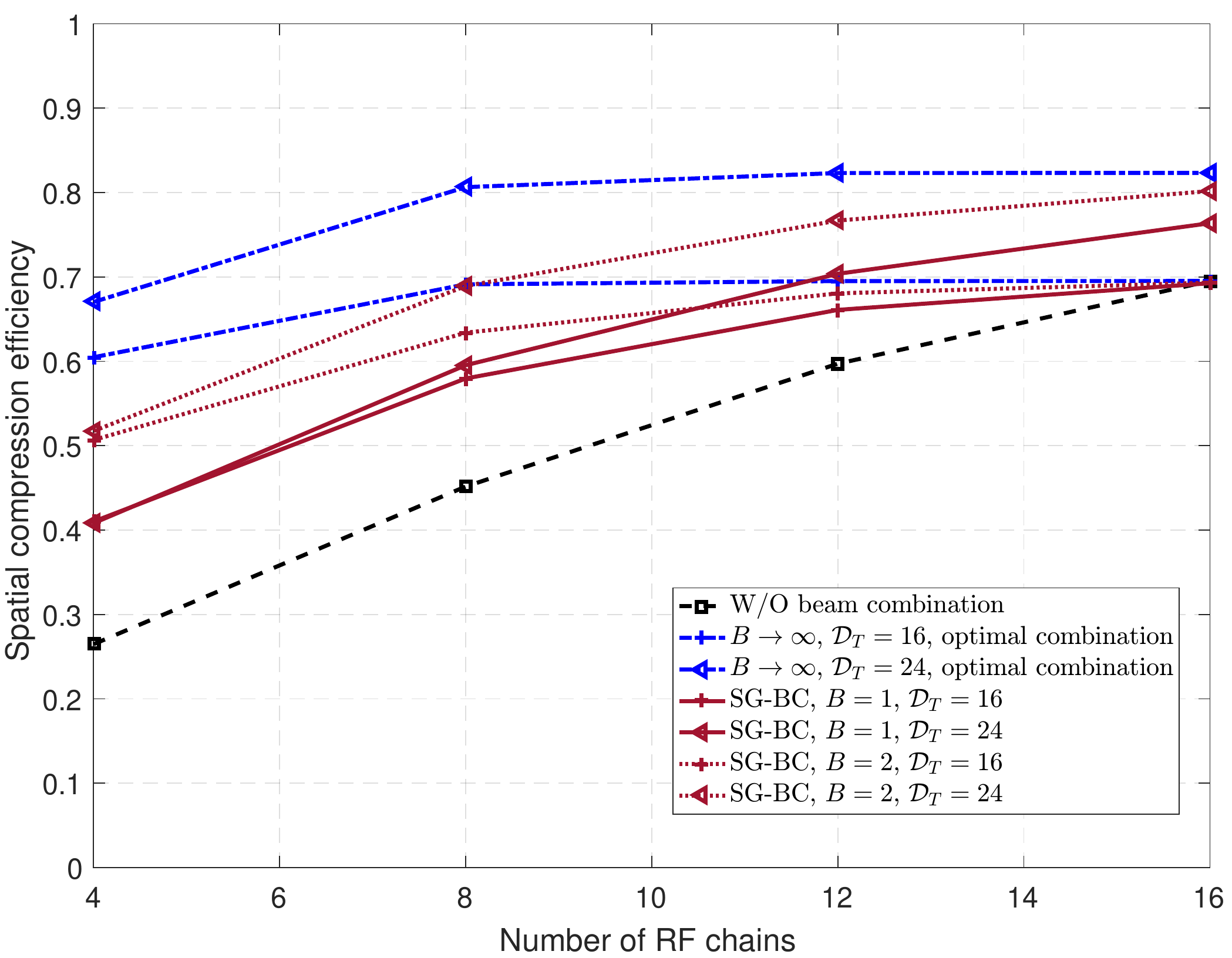}
	\caption{Comparisons of spatial compression schemes with $128$ antenna ports and various number of RF chains. The number of users is $2$, and the user moving speed is $3$~km/h. The UL SNR is $0$~dB.}
	\label{Fig_pr_A128_vB_Bd16_v3_n2}
\end{figure}

\subsection{How Many RF Chains Can Be Saved by the Proposed Beam Combination Schemes?}
To answer the question of how many RF chains can be saved and meanwhile achieving the same compression efficiency, we investigate the impact of the number of RF chains after beam combination on different spatial compression schemes. In Fig. \ref{Fig_pr_A128_vB_Bd16_v3_n2}, the parameter setting is the same as in Fig. \ref{Fig_pr_A128_B8_Bd16_v3_n2}. It is observed that about $4$ RF chains can be saved by adopting a PSN with $24$ $2$-bit digital phase shifters. 

\begin{figure}[!t]
	\centering
	\includegraphics[width=0.45\textwidth]{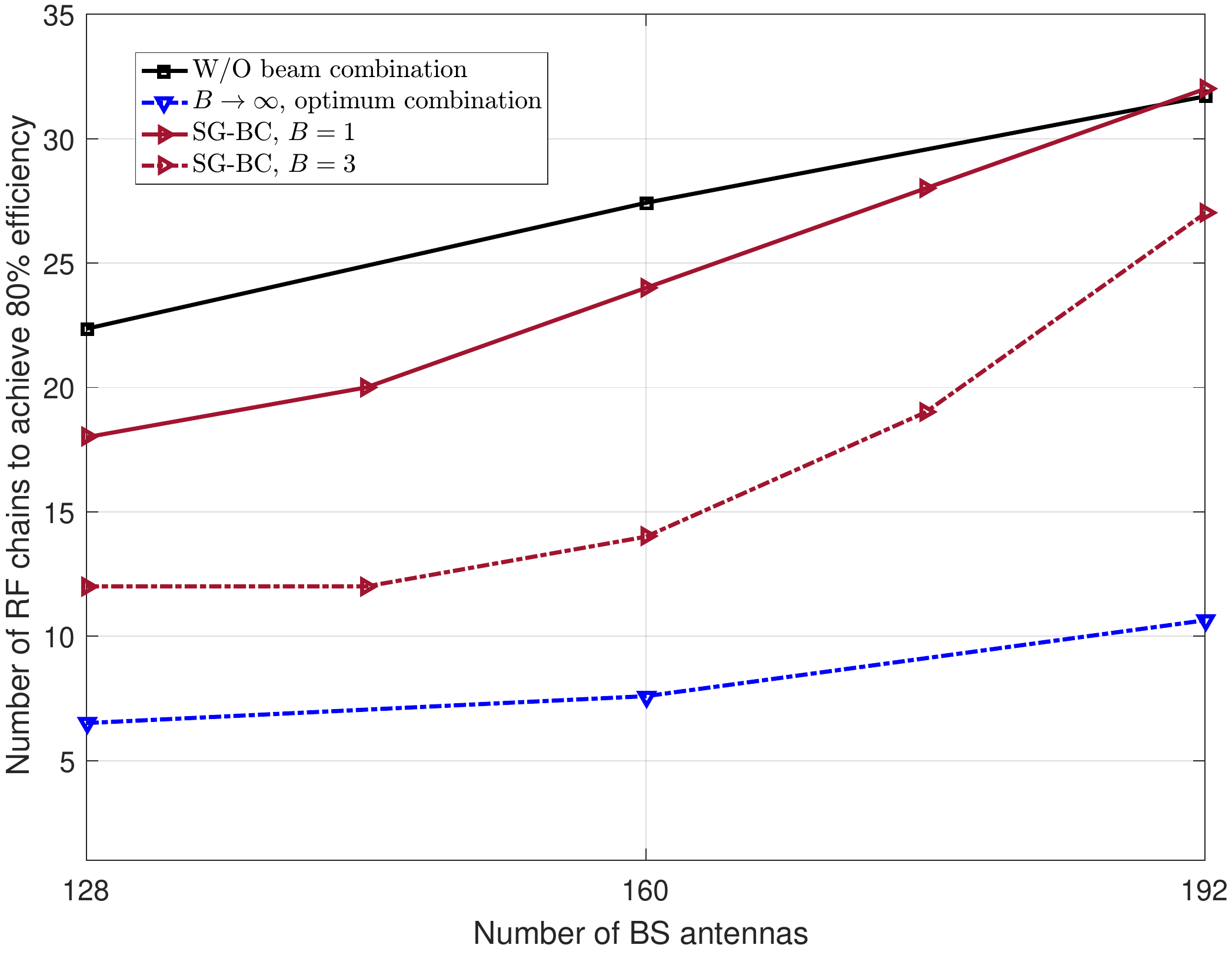}
	\caption{Comparisons of spatial compression schemes to achieve $80\%$ spatial compression efficiency. The number of beamspace beams is $32$. The number of users is $2$, and the user moving speed is $3$~km/h. The UL SNR is $0$~dB.}
	\label{Fig_B_vA_80pr_v3_n2}
\end{figure}

In Fig. \ref{Fig_B_vA_80pr_v3_n2}, the number of RF chains that are sufficient to attain $80\%$ spatial compression efficiency is investigated. Significant RF chain reduction is possible based on the proposed spatial compression schemes, e.g., with $160$ BS antennas, a PSN with $32$ $3$-bit phase shifters can reduce the number of RF chains from $27$ to $15$, while maintaining most of the signal power. Even with one-bit PSN, about $4$ RF chains can be saved with $128$ BS antennas.

\subsection{Impact of Phase Shifter Resolution and Signal Angular Spread}

\begin{figure}[!t]
	\centering
	\includegraphics[width=0.45\textwidth]{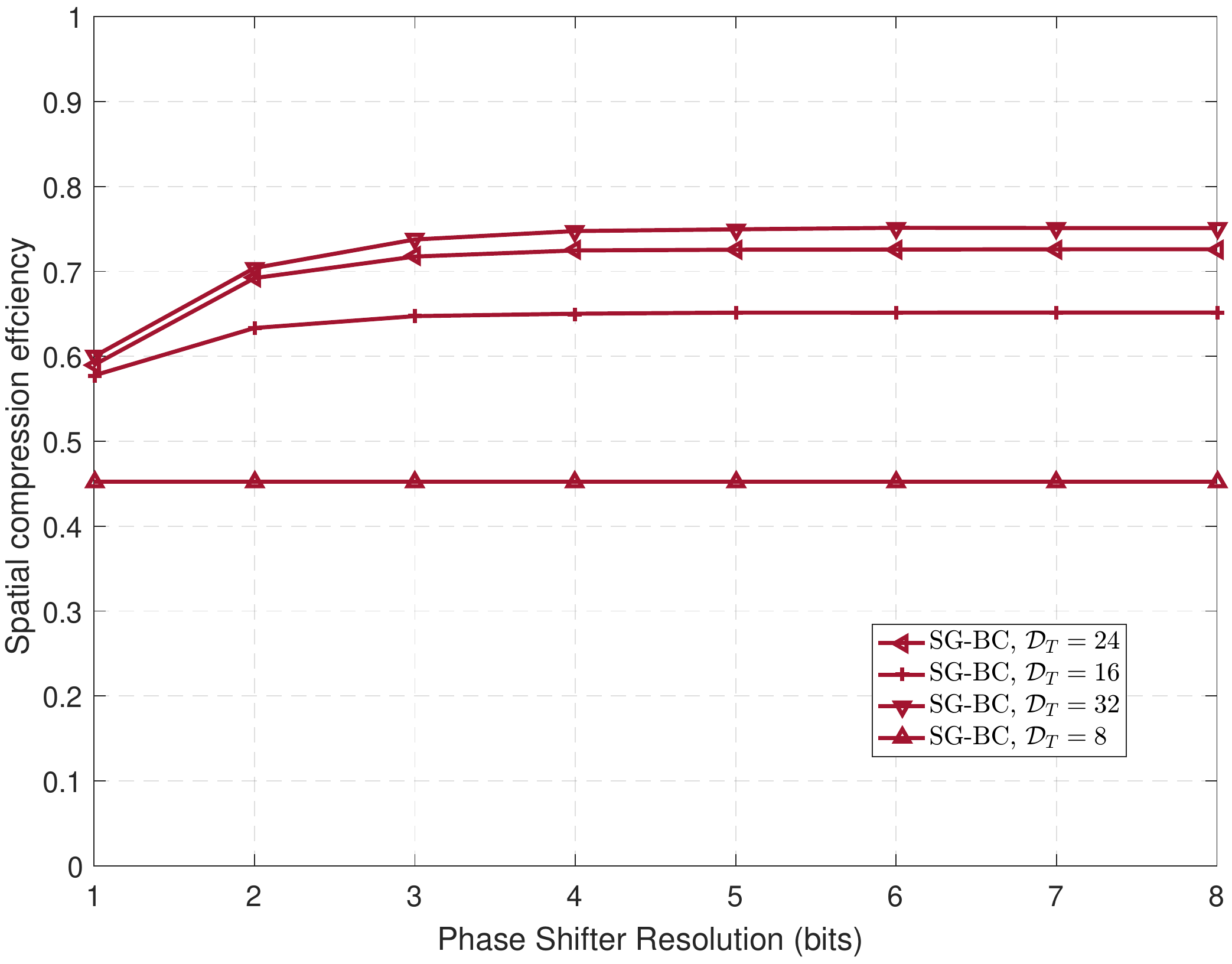}
	\caption{The impact of phase shifter resolutions. The number of RF chains is $8$. The number of users is $2$, and the user moving speed is $3$~km/h. The UL SNR is $0$~dB.}
	\label{Fig_res}
\end{figure}
Fig. \ref{Fig_res} demonstrates the impact of PSN resolutions on the system performance. It is shown that a low-resolution PSN (ont-bit and 2-bit PSN) is sufficient since a high-resolution PSN brings marginal performance gain. Note that since $8$ RF chains are used out of $\mathcal{D}_\textrm{T}=8$ beams in the bottom plot, there is no gain in using a higher-cost PSN in this case. Note that even with the resolution going to infinity, there is still performance gap between the PSN and the optimal combination in Theorem \ref{thm3}, due to constant amplitude constraints of the PSN.

\begin{figure}[!t]
	\centering
	\includegraphics[width=0.45\textwidth]{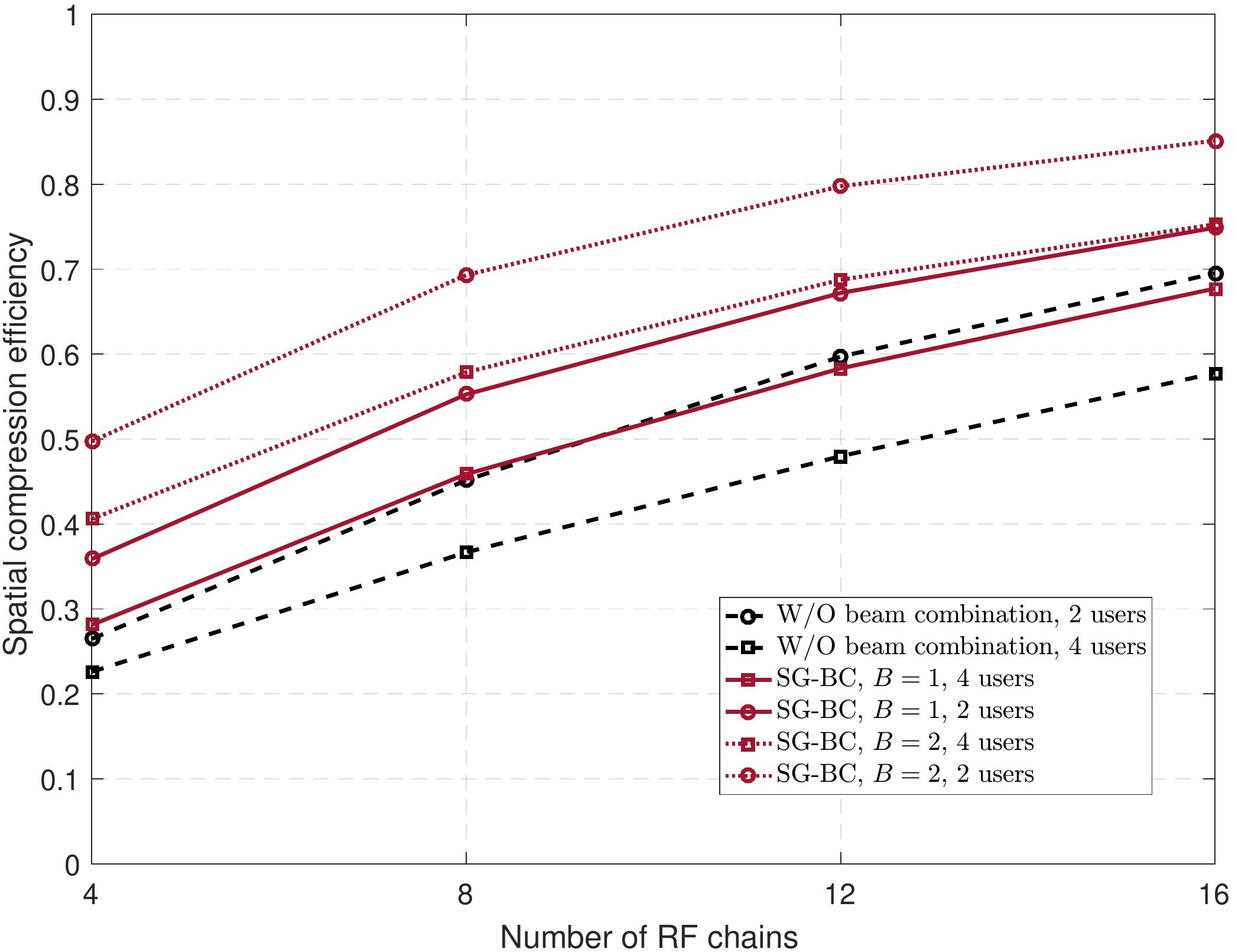}
	\caption{The impact of angular spread on the transform efficiency. The user moving speed is $3$~km/h. The UL SNR is $0$~dB. }
	\label{Fig_vn}
\end{figure}
In Fig. \ref{Fig_vn}, the impact of angular spread of the uplink receive signal is investigated. The total angular spread of the uplink signal is determined by the propagation environment, user locations, and the number of users. In Fig. \ref{Fig_vn}, we compare the number of users of $2$ and $4$ to obtain different angular spread. The user locations are uniformly distributed in the angular domain. Obviously, the $4$-user case has larger angular spread. It is observed that a larger angular spread leads to lower compression efficiency due to the fact that more beams are needed to cover the angular domain. 

It is worthwhile to mention that in a densely deployed cell where users are randomly distributed, the combined signal angular spread of all users is large, and hence the spatial compression gain of the proposed scheme is inevitably reduced. However, considering the millimeter-wave based system where the number of MPCs inside the angular spread is small, the proposed scheme can still provide considerable gain even with a large combined angular spread since the gain is directly related to the number of MPCs.  

\subsection{Link-Level Simulations for Achievable Rates}
\begin{figure}[!t]
	\centering
	\includegraphics[width=0.45\textwidth]{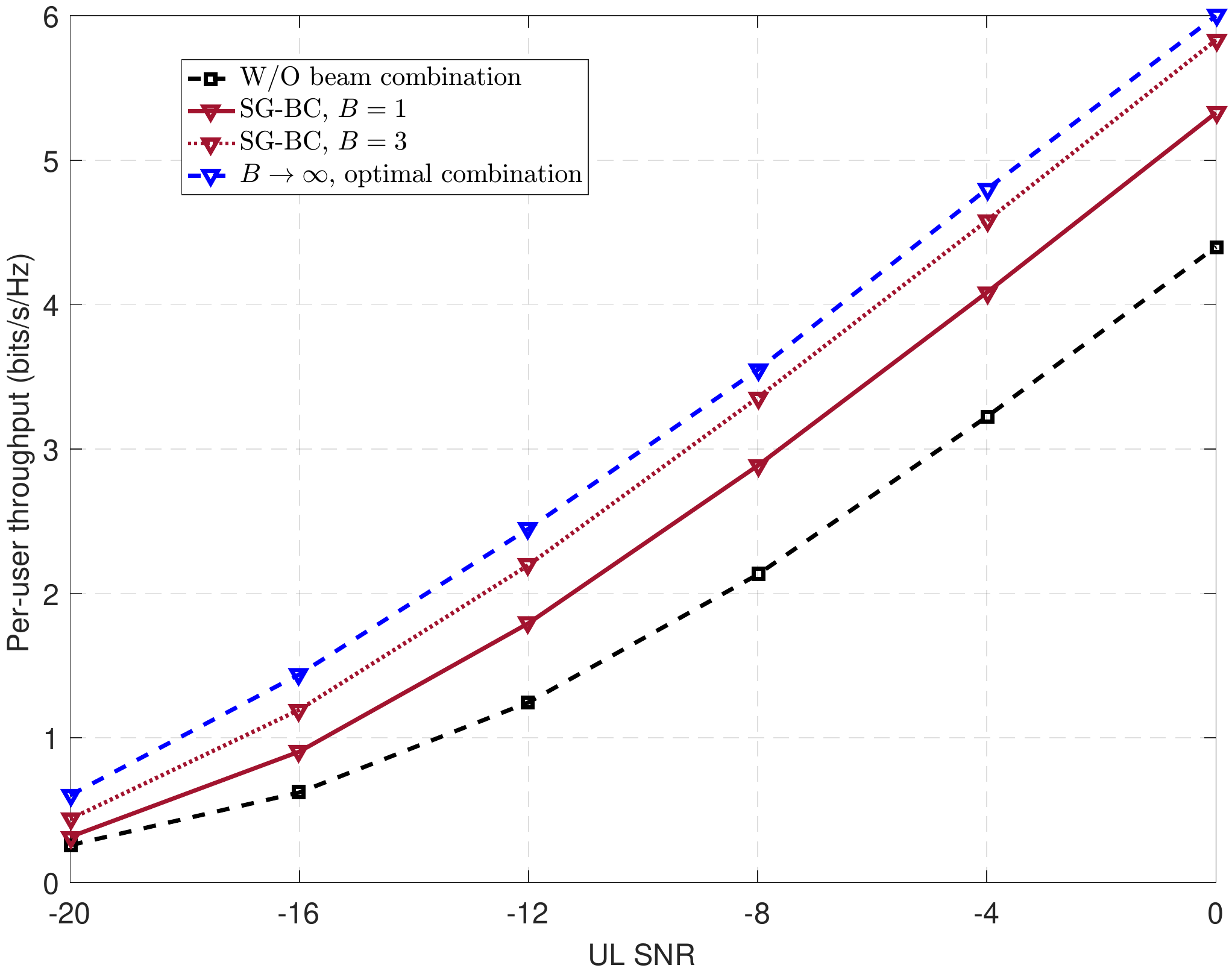}
	\caption{Link-level per-user throughput comparisons with $32$ phase shifters and $8$ RF chains. The number of users is $2$ and the moving speed is $3$~km/h.}
	\label{Fig_A128_B8_Bd16_SNR}
\end{figure}
\begin{figure}[!t]
	\centering
	\includegraphics[width=0.45\textwidth]{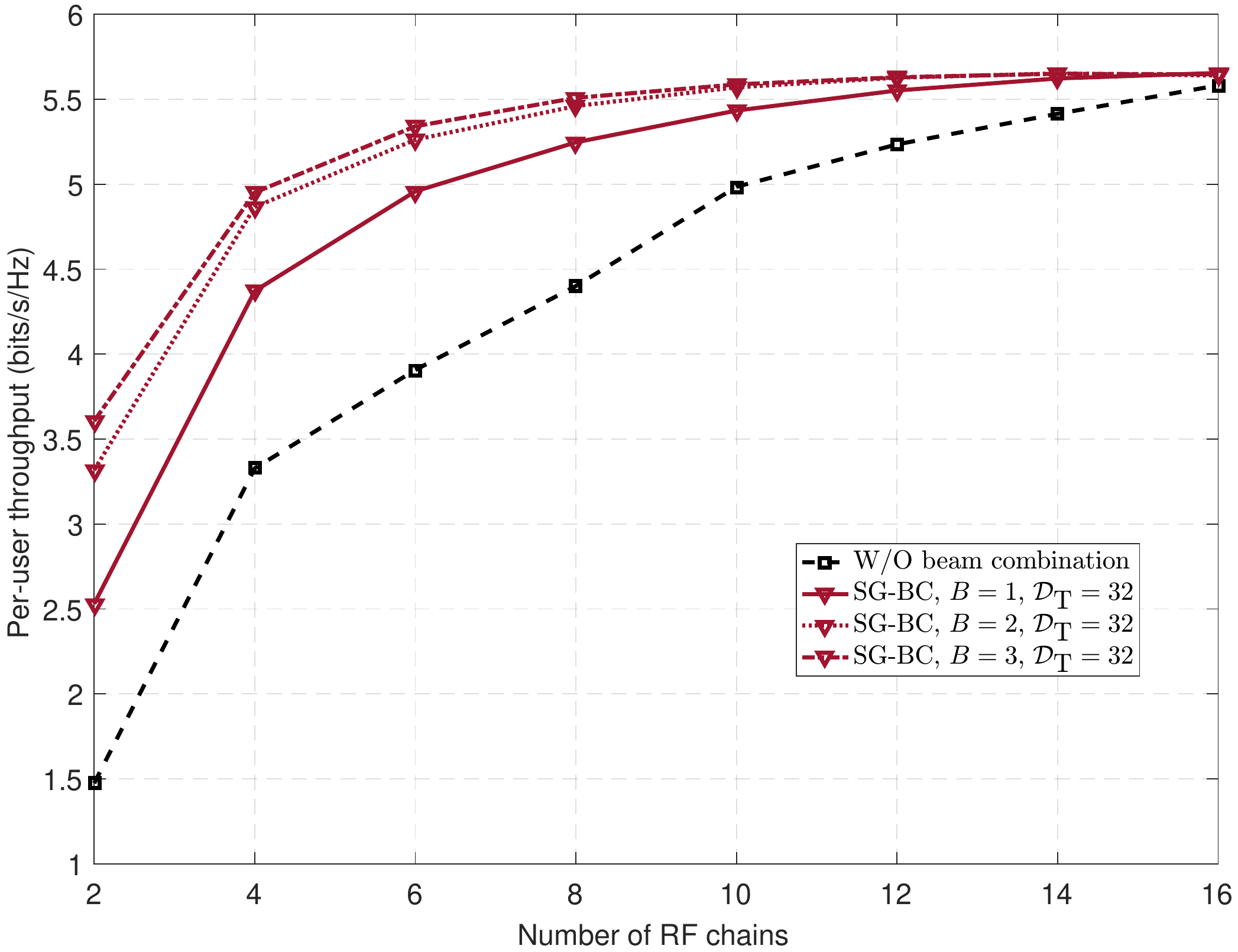}
	\caption{Link-level per-user throughput comparisons with $32$ phase shifters and different number of RF chains. The UL SNR is $0$~dB. The number of users is $2$ and the moving speed is $3$~km/h.}
	\label{Fig_A128_Bd16_xB}
\end{figure}
In order to validate the proposed spatial compression performance in practice and also show that the compression efficiency metric is well related to real-system performance, a link-level LTE-based simulation is conducted. The spatial compression is performed before the channel estimation module, which adopts a FFT-based scheme \cite{tan06}, and the baseband receiving algorithm to decode multi-user signals is MMSE-based. After the MMSE receiver, the decoded constellation points for the user are compared with the transmit ones to calculate the symbol-error-rate (SER). The simulator does not include channel coding and decoding to save processing time. The candidate modulation schemes are quadrature phase-shift keying (QPSK), $16$-quadrature-amplitude-modulation ($16$-QAM) and $64$-QAM. The SINR is mapped from the SER based on a predefined look-up table (with different modulation orders) and thereby the throughput is calculated based on the Shannon formula with the SINR derived before. The simulator only calculates the throughput of the first user for simplicity, and averaged over multiple drops. Therefore, the resulting throughput can be interpreted as per-user throughput. The link adaptation is enabled to support various SNRs whereby the BS estimates the SINR based on received sounding-reference-signals (SRSs) to decide the uplink transmit modulation-coding-scheme (MCS). No outer-loop link adaptation is used. All users are scheduled simultaneously (traffic type: full buffer) on the whole frequency bandwidth. The SRS which follows the 3GPP definition of Zadoff-Chu (ZC) sequences \cite{zep13} is enabled to simulate the LTE-based uplink traffic channel (PUSCH) transmissions. We run each drop for $200$~{ms}, which corresponds to $20$ radio frames in the LTE systems.

Fig. \ref{Fig_A128_B8_Bd16_SNR} shows the comparison for a typical scenario, where the BS has $128$ antenna ports, i.e., $2 \times 32 \times 2$ (rows $\times$ columns $\times$ polarizations), and the number of RF chains is $8$. The number of beams after beamspace transformation is varied from $8$ to $32$. Similar performance trend as in Fig. \ref{Fig_A128_B8_Bd16_SNR} is observed, which shows that the proposed beam combination schemes can achieve higher throughput than the conventional beamspace MIMO system without beam combination given the same number of RF chains.

In Fig. \ref{Fig_A128_Bd16_xB}, the effect of RF chain reduction is presented by throughput simulation results. It is observed that considerable RF complexity reduction is possible by the proposed spatial compression schemes.

\section{Conclusions and Discussions}
\label{sec:conc}
In this paper, we propose to adopt a spatial compression module after the beamspace transformation in lens antenna array to further reduce the RF complexity in massive MIMO systems. The fundamental reason that the RF chains can be saved by the proposed beam combination schemes is spatial power leakage by the lens antenna array and imperfect channel statistics estimations. In order to implement the idea with low hardware cost. We propose to realize the spatial compression module with low-resolution constant-amplitude digital phase shifters. The optimal discrete beam combination with the hardware constrains is solved by the BB-BC scheme which is based on the BB methodology. The optimum solution to the sub-problem in the BB-BC is given in a closed-form which is key to the BB scheme. Based on the structure of the optimum solution to the BB sub-problem, a low-complexity SG-BC scheme is proposed whose computational complexity scales linearly with the number of beams. The number of phase shifters in the PSN given the signal angular spread is also derived in a closed-form. 

The spatial compression efficiency is used as the metric to compare proposed schemes and benchmarks based on a 3GPP SCM. It is observed that the proposed spatial compression module can reduce the number of RF chains, and hence the RF complexity with low additional cost. In a typical urban scenario where the BS is equipped with $128$ antennas, the number of RF chains can be cut down up to $25\%$ by a one-bit PSN and to $40\%$ by a $2$-bit PSN with $32$ phase shifters. It is shown that the low-complexity SG-BC scheme performs fairly close to the optimal BB-BC scheme. Based on the proposed spatial compression scheme, the phase shifter resolution does not need to be high ($2$ bit, even one-bit, is sufficient in most scenarios). In order to check the compatibility with other processing blocks and system achievable rates, a full link-level signal processing chain based on LTE numerologies and 3GPP SCM is simulated, wherein uplink signals are decoded and the achievable rates are consequently obtained. It validates that the proposed schemes are effective in practice, and thus provides a promising solution for the massive MIMO implementation in $5$G systems.

Regarding future directions, more sophisticated signaling exchange between the BS scheduler and the RF module should be considered. The spatial compression schemes proposed by this paper adopts a signal-power-based criterion to combine beams in the beamspace. In other words, the proposed scheme is solely based on optimizing the received signal power. However, whether this signal is useful signal or interference is not considered. On the other hand, user fairness is also ignored, which means that when user signals are power imbalanced, the stronger signal from some users would drown the weak ones. Although the signal-power-based criterion makes much sense when the RF module is self-contained and do not exchange high-layer control messages with the BS scheduler, these two potential problems are very relevant in practice, and they can be alleviated by well-designed signaling and protocols that worth further investigations.

\appendices
\section{Proof of Theorem \ref{thm3}}
\label{app2}
\begin{IEEEproof}
	Consider the transform efficiency maximization problem,
	\begin{eqnarray}
	\label{P1}
	\max_{\bm{F}} && \eta(\bm{F})\nonumber\\
	\textrm{s.t.,} && \bm{F} \in \mathbb{H}_{N_\textrm{s}},
	\end{eqnarray}
	wherein the objective can be derived as
	\begin{equation}
	\label{eta_p}
	\eta(\bm{F}) = \frac{\textrm{tr}\left[\bm{F}\bar{\bm{R}_\textrm{s}}\bm{F}^\dag\right]}{\textrm{tr}\,\bar{\bm{R_\textrm{s}}}} \approx \frac{\textrm{tr}\left[\bm{F}\bar{\bm{R}_\textrm{t}}\bm{F}^\dag\right]- N_\textrm{s} \sigma^2}{\textrm{tr}\,\bar{\bm{R_\textrm{t}}} - N \sigma^2},
	\end{equation}
	which is equivalent to maximizing 
	\begin{equation}
	\eta^\prime(\bm{F}) = \frac{\textrm{tr}\left[\bm{F}\bar{\bm{R}_\textrm{t}}\bm{F}^\dag\right]}{\textrm{tr}\,\bar{\bm{R_\textrm{t}}}}.
	\end{equation}
	Denote the SVD as $\bar{\bm{R}}_\textrm{t} = \bm{U}\bm{S}\bm{U}^\dag$, and $\bm{G} = \bm{F}\bm{U}$ which is a bijection in $\mathbb{H}_{N_\textrm{s}}\rightarrow\mathbb{H}_{N_\textrm{s}}$. Therefore, it is equivalent to substitute $\bar{\bm{R}}_\textrm{t}$ in \eqref{eta_p} for $\bm{S}$. Denote $\bar{\bm{G}} = \left[\bm{G}^\dag,\,\bm{G}_\perp^\dag\right]^\dag$, where $\bm{G}_\perp$ is a matrix that, together with $\bm{G}$, makes $\bar{\bm{G}}$ a unitary matrix. It follows that
	\begin{IEEEeqnarray}{rCl}
		\eta^\prime(\bm{F}) &=& \frac{\textrm{tr}\left[\bm{G}{\bm{S}}\bm{G}^\dag\right]}{\textrm{tr}\,{\bm{S}}} \nonumber \\
		&=& \frac{\textrm{tr}\left[\bar{\bm{G}}{\bm{S}}\bar{\bm{G}}^\dag \left[\begin{array}{*{20}{c}}
                        {\bm{I}_{{N_\textrm{s}}}}&{}\\
                        {}&\bm{0}
                        \end{array}\right] \right]}{\textrm{tr}\,{\bm{S}}} \nonumber\\
		&\overset{(a)}{\le}& \frac{\textrm{tr}\,{\bm{S}_1}}{\textrm{tr}\,{\bm{S}}},
	\end{IEEEeqnarray}
	where $\bm{S}_1$ and $\bm{S}_2$ are diagonal matrices containing the first $N_\textrm{s}$ and the last $N-N_\textrm{s}$ diagonal elements. The inequality of $(a)$ is based on \cite[Lemma 1]{zhou02}. It is straightforward to observe that the equality holds if $\bm{G}$ consists of the first $N_\textrm{s}$ rows of an $N$-dimensional identity matrix, such that the solution to \eqref{P1} is the first $N_\textrm{s}$ ($N_\textrm{s} \le N$) columns of the singular matrix of the CCM. Combining with \eqref{eta_p},
	\begin{equation}
	\eta(\bm{F_\textrm{opt}}) = \frac{\textrm{tr}\,{\bm{S}_1} - N_\textrm{s} \sigma^2}{\textrm{tr}\,{\bm{S}}- N \sigma^2} = \frac{\sum_i^{N_\textrm{s}}\left({\lambda_i}-\sigma^2 \right) }{\sum_i^{N}\left({\lambda_i} - \sigma^2 \right)}.
	\end{equation}
	Since in most cases (medium to high SNR) the approximation in (\ref{eta_p}) is accurate, $\bm{F}_{\textrm{opt}}$ can also maximize $\eta$, despite the fact that it is derived based on maximizing $\eta^\prime$.
\end{IEEEproof}

\section{Proof of Proposition \ref{thm2}}
\label{pr_th1}
\begin{IEEEproof}
	Consider the beamspace transformation in \eqref{array_res} and \eqref{beamspace_tr}. Denote 
	\begin{IEEEeqnarray}{rcl}
		\bm{e}(\sin(\theta_i)) &=& \frac{1}{\sqrt{M}} \left[1, \exp\left(-\frac{j2\pi d \sin(\theta_i)}{\lambda}\right),...,\right. \nonumber\\
		&& \left.\exp\left(-\frac{j2\pi (M-1)d \sin(\theta_i)}{\lambda}\right) \right]^\dag.
	\end{IEEEeqnarray}
	Denote the beamspace channel vector as
	\begin{equation}
	\label{DFT}
	\bm{h}^\textrm{a} = \bm{A}_\mathcal{L} \bm{h},
	\end{equation}
	then,
	\begin{IEEEeqnarray}{rcl}
		h^\textrm{a}_m &=& \sum_{i=1}^{B} a_i e^{j\phi} \omega_{m,i}, \nonumber\\
		\omega_{m,i} &=&  \bm{e}^\dag\left(\frac{m-1}{Md/\lambda}\right)\bm{e}(\sin(\theta_i)),
	\end{IEEEeqnarray}
	and
	\begin{equation}
	\omega_{m,i} = \frac{1}{M}\exp(j\pi (M-1)dC_{m,i}/\lambda)\frac{\sin(\pi Md C_{m,i}/\lambda)}{\sin(\pi Md C_{m,i} /(M\lambda))},
	\end{equation}
	where $C_{m,i} = \frac{m-1}{Md/\lambda}-\sin(\theta_i)$. For
	\begin{equation}
	\label{deltacos}
	|C_{m,i}|>\frac{\sqrt{M}\lambda}{Md}=\frac{1}{\sqrt{M}d},
	\end{equation}
	it follows that,
	\begin{IEEEeqnarray}{rcl}
		\label{innerp}
		|\omega_{m,i}|&\le&\frac{1}{M |\sin(\pi Md C_{m,i} /(M\lambda))|} \nonumber\\
		&<&\frac{1}{M\sin(\pi/\sqrt{M})} \nonumber \\
		&\stackrel{M \to \infty}{\longrightarrow}& 0.
	\end{IEEEeqnarray}
	Combining \eqref{deltacos} and \eqref{innerp}, it is proved that in the large system limit, the $m$-th component of the angular representation, $h^\textrm{a}_m$ is \emph{non-zero} up to a constant representing the side lobes shown in \eqref{deltacos}, only if $\frac{m-1}{Md/\lambda}$ is within the angular spread of the MPC directional sines. Therefore, given the interval between DFT base vectors as $\frac{1}{Md/\lambda}$, the number of non-zero components in $\bm{h}^\textrm{a}$, denoted by $\gamma$, is
	\begin{equation}
	\gamma = \frac{\Omega}{\frac{1}{Md/\lambda}} + \frac{2}{\sqrt{M}d} \stackrel{M \to \infty}{\longrightarrow} \frac{Md}{\lambda}\Omega.
	\end{equation}
	
	Consider the angular representation of the downlink channel matrix $\bm{H}$,
	\begin{equation}
	\label{Ha}
	\bm{H}^\textrm{a} =  \bm{H} \bm{A}_\mathcal{L}^\dag = \left[\bm{h}^\textrm{a}_1,\bm{h}^\textrm{a}_2,...,\bm{h}^\textrm{a}_N\right]^\dag.
	\end{equation}
	Notice that the $i$-th column of $\bm{H}^\textrm{a}$ is non-empty iff. there exists one user with angular spread measured in directional sine overlapping with $\frac{i-1}{Md/\lambda}$. Therefore, the total number of non-empty columns is the union of the angular spread of all users measured in directional sine, up to a multiplicative constant. This concludes the proof.
\end{IEEEproof}

\section{Proof for Theorem \ref{thm_opt}}
\label{app1}
The objective function value is 
\begin{IEEEeqnarray}{rCl}
&& \eta(\bm{w}_\textrm{J}) \nonumber\\
\label{eta_w}
&=& \frac{\bm{d}_\textrm{I}^\dag \bm{R}_\textrm{I} \bm{d}_\textrm{I}+\bm{w}_\textrm{J}^\dag \bm{R}_\textrm{J}\bm{w}_\textrm{J} + \bm{p}^\dag \bm{w}_\textrm{J} + \bm{w}_\textrm{J}^\dag \bm{p}}{\bm{d}_\textrm{I}^\dag \bm{d}_\textrm{I}+\bm{w}_\textrm{J}^\dag \bm{w}_\textrm{J}} \\
&=& \frac{\bm{d}_\textrm{I}^\dag \bm{R}_\textrm{I} \bm{d}_\textrm{I}+\bm{w}_\textrm{J}^\dag \bm{R}_\textrm{J}\bm{w}_\textrm{J} + \bm{p}^\dag \bm{G}^{-\frac{1}{2}} \bm{G}^{\frac{1}{2}} \bm{w}_\textrm{J} + \bm{w}_\textrm{J}^\dag \bm{G}^{\frac{1}{2}} \bm{G}^{-\frac{1}{2}} \bm{p}}{\bm{d}_\textrm{I}^\dag \bm{d}_\textrm{I}+\bm{w}_\textrm{J}^\dag \bm{w}_\textrm{J}} \nonumber\\
\label{ineq1}
&\le& \frac{\bm{d}_\textrm{I}^\dag \bm{R}_\textrm{I} \bm{d}_\textrm{I}+\bm{w}_\textrm{J}^\dag \bm{R}_\textrm{J}\bm{w}_\textrm{J} + \bm{w}_\textrm{J}^\dag \bm{G} \bm{w}_\textrm{J} + \bm{p}^\dag \bm{G}^{-1} \bm{p}}{\bm{d}_\textrm{I}^\dag \bm{d}_\textrm{I}+\bm{w}_\textrm{J}^\dag \bm{w}_\textrm{J}} \\
&=& \frac{\bm{d}_\textrm{I}^\dag \left(\bm{R}_\textrm{I} + \bm{R}_\textrm{JI}^\dag \bm{G}^{-1} \bm{R}_\textrm{JI} \right) \bm{d}_\textrm{I}+\bm{w}_\textrm{J}^\dag \left(\bm{R}_\textrm{J} + \bm{G}\right) \bm{w}_\textrm{J}}{\bm{d}_\textrm{I}^\dag \bm{d}_\textrm{I}+\bm{w}_\textrm{J}^\dag \bm{w}_\textrm{J}} \nonumber\\
\label{ineq2}
&\le& \frac{\bm{d}_\textrm{I}^\dag \left(\bm{R}_\textrm{I} + \bm{R}_\textrm{JI}^\dag \bm{G}^{-1} \bm{R}_\textrm{JI} \right) \bm{d}_\textrm{I}+ \tilde{\lambda} \bm{w}_\textrm{J}^\dag \bm{w}_\textrm{J}}{\bm{d}_\textrm{I}^\dag \bm{d}_\textrm{I}+\bm{w}_\textrm{J}^\dag \bm{w}_\textrm{J}},
\end{IEEEeqnarray}
where $\bm{G}$ is a positive semi-definite matrix. The inequality in \eqref{ineq1} stems from the fact that 
\begin{IEEEeqnarray}{rCl}
&& \left\|\bm{G}^{\frac{1}{2}} \bm{w}_\textrm{J} - \bm{G}^{-\frac{1}{2}} \bm{p} \right\|_2^2 \nonumber\\
&=& \bm{w}_\textrm{J}^\dag \bm{G} \bm{w}_\textrm{J} + \bm{p}^\dag \bm{G}^{-1} \bm{p} - \bm{p}^\dag \bm{G}^{-\frac{1}{2}} \bm{G}^{\frac{1}{2}} \bm{w}_\textrm{J} - \bm{w}_\textrm{J}^\dag \bm{G}^{\frac{1}{2}} \bm{G}^{-\frac{1}{2}} \bm{p} \ge 0.\nonumber\\
\end{IEEEeqnarray}
The equality is upheld if and only if 
\begin{equation}
\label{eq1}
\bm{G} \bm{w}_\textrm{J} = \bm{p}.
\end{equation}
Note that the relationship between $\bm{w}_\textrm{J}$ and $\bm{p}$ in \eqref{eq1} is without loss of generality for the maximization problem since the only constraint it introduces is 
\begin{equation}
\bm{w}_\textrm{J}^\dag \bm{p} = \bm{w}_\textrm{J}^\dag \bm{G} \bm{w}_\textrm{J} \ge 0.
\end{equation}
Additionally, given the objective function in \eqref{eta_w}, such a constraint is reasonable since $\forall \bm{w}_{\textrm{J},0}$ satisfying $\bm{w}_{\textrm{J},0}^\dag \bm{p} \le 0$, $\eta(\bm{w}_{\textrm{J},0}) \le \eta(-\bm{w}_{\textrm{J},0})$. However, it is problematic when $\bm{w}_\textrm{J}=\bm{0}$ or $\|\bm{w}_\textrm{J}\|_2 \to \infty$ in \eqref{eq1}, and hence both circumstances will be dealt with separately later.

The inequality in \eqref{ineq2} follows from the definition of the Euclidean norm of a positive semi-definite matrix $\bm{R}_\textrm{J} + \bm{G}$, where $\tilde{\lambda}$ is the largest singular value of $\bm{R}_\textrm{J} + \bm{G}$, and 
\begin{equation}
\label{eq2}
\left(\bm{R}_\textrm{J}+\bm{G}\right) \bm{w}_\textrm{J} = \tilde{\lambda} \bm{w}_\textrm{J},
\end{equation}
Letting
\begin{equation}
\label{eq3}
\tilde{\lambda} = \frac{\bm{d}_\textrm{I}^\dag \left(\bm{R}_\textrm{I} + \bm{R}_\textrm{JI}^\dag \bm{G}^{-1} \bm{R}_\textrm{JI} \right) \bm{d}_\textrm{I}}{\bm{d}_\textrm{I}^\dag \bm{d}_\textrm{I}},
\end{equation}
we can obtain
\begin{equation}
\eta(\bm{w}_\textrm{J}) \le \tilde{\lambda}.
\end{equation}
The task now is to find the optimum value and solution to the problem given the equality equations of \eqref{eq1}, \eqref{eq2} and \eqref{eq3}. Plugging \eqref{eq1} into \eqref{eq2}, it follows that
\begin{equation}
\label{bb}
\bm{R}_\textrm{J} \bm{w}_\textrm{J} + \bm{p}= \tilde{\lambda} \bm{w}_\textrm{J}.
\end{equation}

For mathematical rigour, two \emph{limiting cases} should be treated separately, i.e.,
\begin{equation}
\label{lim}
\bm{w}_\textrm{J}=\bm{0}\textrm{, or }\|\bm{w}_\textrm{J}\|_2 \to \infty.
\end{equation}
When $\bm{w}_\textrm{J}=\bm{0}$, $\eta(\bm{w}_\textrm{J})=r/d$. When $\|\bm{w}_\textrm{J}\|_2 \to \infty$, it follows that  
\begin{equation}
    \eta(\bm{w}_\textrm{J}) \to \frac{\bm{w}_\textrm{J}^\dag \bm{R}_\textrm{J}\bm{w}_\textrm{J}}{\bm{w}_\textrm{J}^\dag \bm{w}_\textrm{J}} \le \lambda_1.
\end{equation}
and the optimum 
\begin{equation}
\label{o1}
    \bm{w}_{\textrm{J},\infty}^{*} = \beta \bm{u}_\textrm{J,dom},
\end{equation}
where $\bm{u}_\textrm{J,dom}$ is a unit-norm dominant singular vector of $\bm{R}_\textrm{J}$ and $\beta \to \infty$.

Having disposed of the limiting cases, we can now proceed. Since by adopting \eqref{o1} and $\bm{w}_\textrm{J}=\bm{0}$, it yields the objective function value of $\lambda_1$ and $r/d$ respectively, it can be concluded that $\lambda^* > \max[\lambda_1,r/d]$. Stemming from \eqref{bb}, we can obtain
\begin{equation}
\label{o2}
\bm{w}_\textrm{J} = \left(\tilde{\lambda} \bm{I}_{L-l}- \bm{R}_\textrm{J} \right)^{-1} \bm{p}.
\end{equation}
Since $\lambda^* > \lambda_1$, the matrix $\tilde{\lambda} \bm{I}_{L-l}- \bm{R}_\textrm{J}$ is always invertible. Plugging \eqref{o2} into \eqref{eq3} and using the denotations in \textbf{P1}, we can obtain
\begin{IEEEeqnarray}{rCl}
	\label{o3}
\tilde{\lambda} d - r &=& \bm{p}^\dag \left(\tilde{\lambda} \bm{I}_{L-l}- \bm{R}_\textrm{J} \right)^{-1} \bm{p} \\
&=& \bm{p}^\dag \bm{U}_\textrm{J} \left(\tilde{\lambda} \bm{I}_{L-l}- \bm{\Sigma}_\textrm{J} \right)^{-1}  \bm{U}_\textrm{J}^\dag \bm{p} \\
&=& \sum_{i=1}^{L-l} \frac{\left|\left(\bm{U}_\textrm{J}^\dagger \bm{p}\right)_i\right|^2}{\tilde{\lambda}-\lambda_i}.
\end{IEEEeqnarray}
Solving \eqref{o3} will give us the optimum value $\lambda^*$. The optimal solution $\bm{w}_\textrm{J}^*$ is given by \eqref{o2}. However, since there are more than one solution to the equation in \eqref{o3}, the problem is which one is the optimum. We proceed to show that the optimum value
\begin{equation}
\label{interval}
\lambda^* \in \left(\lambda_1,\,\frac{d\lambda_1+r+\sqrt{(d\lambda_1-r)^2+4d\bm{p}^\dagger\bm{p}}}{2d}\right],
\end{equation}
and that there is a unique solution to the equation in \eqref{o3} in this interval. 

Concretely, it has been already obtained that $\lambda^*>\lambda_1$. Define
\begin{equation}
\label{ge0}
f(\tilde{\lambda}) = \tilde{\lambda} d - r - \sum_{i=1}^{L-l} \frac{\left|\left(\bm{U}_\textrm{J}^\dagger \bm{p}\right)_i\right|^2}{\tilde{\lambda}-\lambda_i}.
\end{equation}
It is straightforward that $f(\tilde{\lambda})$ is monotonically increasing in the interval of $(\lambda_1,\,\infty)$. 

If there exists some dominant singular vector $\bm{u}_\textrm{J,dom}$ of $\bm{R}_\textrm{J}$ that satisfies $\bm{u}_\textrm{J,dom}^\dag \bm{p} \neq 0$, we obtain 
\begin{equation}
\lim_{\tilde{\lambda} \to \lambda_1^+} f(\tilde{\lambda}) \to -\infty,\textrm{ and } \lim_{\tilde{\lambda} \to +\infty} f(\tilde{\lambda}) \to +\infty,
\end{equation}
there must be a unique value $\lambda^*$ which yields $f(\lambda^*) = 0$ when $\lambda^*>\lambda_1$. Moreover, $\lambda^*>r/d$ based on \eqref{ge0}. The upper bound can be obtained by solving the inequality
\begin{equation}
\sum_{i=1}^{L-l} \frac{\left|\left(\bm{U}_\textrm{J}^\dagger \bm{p}\right)_i\right|^2}{\tilde{\lambda}-\lambda_i} = \tilde{\lambda} d - r \le \frac{\bm{p}^\dag\bm{p}}{\tilde{\lambda} - \lambda_1}.
\end{equation}

On the other hand, if $\forall \bm{u}_\textrm{J,dom}$ we have $\bm{u}_\textrm{J,dom}^\dag \bm{p} = 0$, then it is unclear whether $\lim_{\tilde{\lambda} \to \lambda_1^+} f(\tilde{\lambda}) \to -\infty$ and hence there may not exist $\tilde{\lambda} \in (\lambda_1,+\infty)$ such that $f(\tilde{\lambda})=0$. Specifically, if C1 \eqref{condition1} is not upheld, then there still exists a unique solution of $f(\tilde{\lambda})= 0$ in the interval \eqref{interval}. Otherwise if C1 is satisfied, it follows that the optimum solution is one of the limiting cases discussed in \eqref{lim}, i.e.,
\begin{equation}
    f(\lambda^*) = \max[\lambda_1,r/d].
\end{equation}
With this, we conclude the proof.

\section{The proof of Corollary \ref{coro1}}
\label{app3}
Consider the objective in maximizing
\begin{IEEEeqnarray}{rCl}
	&& \eta(\bm{w}_\textrm{J}) = \frac{\bm{d}_\textrm{I}^\dag \bm{R}_\textrm{I} \bm{d}_\textrm{I}+\bm{w}_\textrm{J}^\dag \bm{R}_\textrm{J}\bm{w}_\textrm{J} + \bm{p}^\dag \bm{w}_\textrm{J} + \bm{w}_\textrm{J}^\dag \bm{p}}{\bm{d}_\textrm{I}^\dag \bm{d}_\textrm{I}+\bm{w}_\textrm{J}^\dag \bm{w}_\textrm{J}} 
\end{IEEEeqnarray}
in the derivation of Theorem \ref{thm_opt} which finds the optimum solution of $\bm{w}_\textrm{J}^*$. An approximation of the optimum solution with discrete constrains in \eqref{ac_const} is one that quantizes $\bm{w}_\textrm{J}^*$. Assume that $\bm{w}_\textrm{J}^*$ obeys complex Gaussian distribution, which is justified if we the channel coefficients follow Rayleigh distributions. Then the quantization error $\bm{e}$ of the quantized beam combination matrix, which is essentially the beam combination matrix with discrete constraints, is given by the rate-distortion theory \cite{Cover12}
\begin{IEEEeqnarray}{rCl}
\bar{\bm{w}}_\textrm{J}^* &=& \bm{w}_\textrm{J}^* + \bm{e},\nonumber\\
\sigma_{\bm{e}}^2 \triangleq \mathbb{E}\left[\bm{e}^\dag \bm{e}\right] &=&  2^{-B} (\bm{w}_\textrm{J}^*)^\dag \bm{w}_\textrm{J}^*
\end{IEEEeqnarray}
where $B$ is the resolution of the PSN. The result is the famous entropy-constrained scalar quantization for Gaussian distributed vectors. Then
\begin{IEEEeqnarray}{rCl}
	&& \eta(\bar{\bm{w}}_\textrm{J}^*) \nonumber\\
	&=& \frac{\bm{d}_\textrm{I}^\dag \bm{R}_\textrm{I} \bm{d}_\textrm{I}+ (\bar{\bm{w}}_\textrm{J}^*)^\dag \bm{R}_\textrm{J} \bar{\bm{w}}_\textrm{J}^* + \bm{p}^\dag \bar{\bm{w}}_\textrm{J}^* + (\bar{\bm{w}}_\textrm{J}^*)^\dag \bm{p}}{\bm{d}_\textrm{I}^\dag \bm{d}_\textrm{I}+(\bar{\bm{w}}_\textrm{J}^*)^\dag \bar{\bm{w}}_\textrm{J}^*}  \nonumber\\
	&=& \frac{\bm{d}_\textrm{I}^\dag \bm{R}_\textrm{I} \bm{d}_\textrm{I}+ (\bm{w}_\textrm{J}^*)^\dag \bm{R}_\textrm{J} {\bm{w}}_\textrm{J}^* + \bm{p}^\dag {\bm{w}}_\textrm{J}^* + ({\bm{w}}_\textrm{J}^*)^\dag \bm{p}}{\bm{d}_\textrm{I}^\dag \bm{d}_\textrm{I}+({\bm{w}}_\textrm{J}^*)^\dag {\bm{w}}_\textrm{J}^* + (\bm{w}_\textrm{J}^*)^\dag \bm{e} + \bm{e}^\dag \bm{w}_\textrm{J}^* + \bm{e}^\dag \bm{e}} \nonumber\\
	&+& \frac{ \bm{e}^\dag \bm{R}_\textrm{J} \bm{e} + \bm{e}^\dag \bm{R}_\textrm{J} {\bm{w}}_\textrm{J}^* + (\bm{w}_\textrm{J}^*)^\dag \bm{R}_\textrm{J} \bm{e} +\bm{p}^\dag \bm{e} + \bm{e}^\dag \bm{p} }{\bm{d}_\textrm{I}^\dag \bm{d}_\textrm{I}+({\bm{w}}_\textrm{J}^*)^\dag {\bm{w}}_\textrm{J}^* + (\bm{w}_\textrm{J}^*)^\dag \bm{e} + \bm{e}^\dag \bm{w}_\textrm{J}^* + \bm{e}^\dag \bm{e}} \nonumber\\
	&\approx& \frac{\bm{d}_\textrm{I}^\dag \bm{R}_\textrm{I} \bm{d}_\textrm{I}+ (\bm{w}_\textrm{J}^*)^\dag \bm{R}_\textrm{J} {\bm{w}}_\textrm{J}^* + \bm{p}^\dag {\bm{w}}_\textrm{J}^* + ({\bm{w}}_\textrm{J}^*)^\dag \bm{p}+ \bm{e}^\dag \bm{R}_\textrm{J} \bm{e} }{\bm{d}_\textrm{I}^\dag \bm{d}_\textrm{I}+({\bm{w}}_\textrm{J}^*)^\dag {\bm{w}}_\textrm{J}^* +  \bm{e}^\dag \bm{e}} \\
	\label{coro1:1}
	&\approx& \frac{\bm{d}_\textrm{I}^\dag \bm{R}_\textrm{I} \bm{d}_\textrm{I}+ (\bm{w}_\textrm{J}^*)^\dag \bm{R}_\textrm{J} {\bm{w}}_\textrm{J}^* + \bm{p}^\dag {\bm{w}}_\textrm{J}^* + ({\bm{w}}_\textrm{J}^*)^\dag \bm{p}+ \sigma_{\bm{e}}^2 \frac{\textrm{tr}\bm{R}_\textrm{J}}{L-l} }{\bm{d}_\textrm{I}^\dag \bm{d}_\textrm{I}+({\bm{w}}_\textrm{J}^*)^\dag {\bm{w}}_\textrm{J}^* +  \sigma_{\bm{e}}^2},\nonumber\\
	\label{coro1:2}
\end{IEEEeqnarray}
where $L-l$ is the dimension of $\bm{w}_\textrm{J}$. The approximation in \eqref{coro1:1} stems from the fact that independent long vectors are asymptotically orthogonal to each other. The approximation in \eqref{coro1:2} is based on \cite[Lemma 14.2]{cou11}.
\bibliographystyle{ieeetr}
\bibliography{12,15}

\end{document}